\documentclass{article}
\usepackage{arxiv}
\newcommand\alumina{$\alpha$-Al\textsubscript{2}O\textsubscript{3}}

\usepackage[strings]{underscore}
\usepackage[utf8]{inputenc} 
\usepackage[T1]{fontenc}    
\usepackage{hyperref}       
\usepackage{url}            
\usepackage{booktabs}       
\usepackage{amsfonts}       
\usepackage{nicefrac}       
\usepackage{microtype}      
\usepackage{graphicx}
\usepackage[numbers,square]{natbib}
\usepackage{appendix}
\usepackage{amssymb}
\usepackage{amsmath,bm}
\usepackage{caption}
\usepackage[capitalise]{cleveref}
\usepackage{csquotes}
\usepackage{subcaption}
\usepackage{adjustbox}
\usepackage{underscore}
\usepackage{multirow}
\usepackage[dvipsnames]{xcolor}
\usepackage{graphicx}
\usepackage{siunitx}
\usepackage{xcolor}
\usepackage[font=small]{caption}
\title{Data-driven inverse uncertainty quantification: application to the Chemical Vapor Deposition Reactor Modeling}

\author{ \href{https://orcid.org/0009-0002-4531-7960}{\includegraphics[scale=0.06]{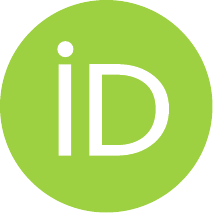}\hspace{1mm}Geremy Loachamin-Suntaxi\thanks{Authors also affiliated with the School of Chemical Engineering, National Technical University of Athens, Zographos Campus, 15780, Attiki, Greece} \thanks{Corresponding author: \texttt{geremy.loachamin@uni.lu}}}\\
	Faculty of Science, Technology and Medicine\\
	University of Luxembourg\\
	Esch-sur-Alzette, L-4364, Luxembourg\\
	\texttt{geremy.loachamin@uni.lu}
	\And
	\href{https://orcid.org/0000-0002-5229-4157}{\includegraphics[scale=0.06]{orcid.pdf}\hspace{1mm}Eleni D.~Koronaki}\\
	Luxembourg Institute of Science and Technology\\
	Esch-sur-Alzette, L-4362, Luxembourg\\
	\texttt{eleni.koronaki@list.lu}
	\And
	\href{https://orcid.org/0000-0003-2272-2584}
	{\includegraphics[scale=0.06]{orcid.pdf}\hspace{1mm}Dimitrios G.~Giovanis} \\
	Department of Civil \& Systems Engineering \\
	Johns Hopkins University\\
	Baltimore, MD 21218, USA\\
	\texttt{dgiovan1@jhu.edu}\\
	\And
	Martin Kathrein \\
	CERATIZIT Luxembourg S.à r.l.\\
	Mamer, L-8232, Luxembourg\\
	\texttt{Martin.Kathrein@ceratizit.com} 
	\And
	Christoph Czettl \\
	CERATIZIT Austria GmbH\\
	Reutte, A-6600, Austria\\
	\texttt{Christoph.Czettl@ceratizit.com} 
	\And
	Andreas G. Boudouvis \\
	School of Chemical Engineering, \\ National Technical University of Athens\\
	Zographos, 15780, Greece\\
	\texttt{boudouvi@chemeng.ntua.gr} 
	\And
	\href{https://orcid.org/0000-0001-7622-2193}{\includegraphics[scale=0.06]{orcid.pdf}\hspace{1mm}St\'{e}phane P.A.~Bordas} \\
	Faculty of Science, Technology and Medicine\\
	University of Luxembourg\\
	Esch-sur-Alzette, L-4364, Luxembourg \\
	\texttt{stephane.bordas@uni.lu}
}

\begin{document}
	\maketitle
	\begin{abstract}
		This study presents a Bayesian framework for (inverse) uncertainty quantification and parameter estimation in a two-step Chemical Vapor Deposition coating process using production data. We develop an XGBoost surrogate model that maps reactor setup parameters to coating thickness measurements, enabling efficient Bayesian analysis while reducing sampling costs. The methodology handles a mixture of data including continuous, discrete integer, binary, and encoded categorical variables. We establish parameter prior distributions through Bayesian Model Selection and perform Inverse Uncertainty Quantification via weighted Approximate Bayesian Computation with summary statistics, providing robust parameter credible intervals while filtering measurement noise across multiple reactor locations. Furthermore, we employ clustering methods guided by geometry embeddings to focus analysis within homogeneous production groups. This integrated approach provides a validated tool for improving industrial process control under uncertainty.
	\end{abstract}
	

	\section{Introduction}\label{sec:Introduction}
	
	Industrial process optimization in complex manufacturing systems presents significant challenges in parameter estimation and (inverse) uncertainty quantification. Chemical vapor deposition (CVD) processes exemplify this complexity, where the intricate interplay between reactor geometry, process conditions, and reactor configurations creates high-dimensional parameter spaces with mixed-type parameters that traditional optimization methods struggle to navigate effectively \cite{koronaki2023partial,gkinis2019building,koronaki2014non,spencer2021investigation}. The need for efficient methodologies capable of inferring physical, operational and set-up parameters from observable coating characteristics motivates the implementation of alternative (data-driven) techniques for (inverse) uncertainty quantification (UQ) \cite{UQ0, UQ1, UQ2}.
	
	Data from industrial processes provide a valuable source for understanding phenomena on a large scale through the implementation of predictive and sampling methods \cite{papavasileiou2024integrating}. However, measurements in CVD processes are inherently expensive and time-consuming to obtain, often prohibiting the comprehensive data collection necessary for robust parameter estimation. These constraints are particularly pronounced due to limited sensor implementation and the substantial resources required for systematic sampling across multiple reactor configurations and operating conditions. Nevertheless, previous studies have demonstrated that it is possible to use actual production data to describe the characteristics of the final product, based on parameters related to CVD reactor configurations \cite{papavasileiou2024integrating, KORONAKI2025109146}.
	
	In Chemical Vapor Deposition (CVD) processes, achieving coating uniformity demands precise control over numerous interdependent parameters, including both numerical and categorical setup variables. While categorical parameters can be encoded through conventional techniques such as binary encoding, recent advances in natural language processing (NLP) offer more powerful alternatives based on embedding representations \cite{KORONAKI2025109146}. However, inherently high-dimensional parameter spaces, which combine heterogeneous variable types, pose substantial challenges for traditional modeling approaches, which often struggle to efficiently capture the statistical descriptors required for robust decision-making and uncertainty quantification.
	
	Traditional methods for uncertainty quantification—such as maximum likelihood estimation and least-squares fitting—provide systematic frameworks for parameter estimation but often lack the flexibility to capture the full extent of uncertainty inherent in complex industrial systems. Bayesian approaches have emerged as powerful alternatives, offering probabilistic formulations that explicitly quantify and propagate uncertainty throughout the inference process \cite{dornheimModelFreeAdaptiveOptimal2020,humfeldMachineLearningFramework2021}. However, conventional Markov Chain Monte Carlo (MCMC) techniques remain computationally demanding and poorly scalable, limiting their applicability in large-scale industrial contexts.
	
	Surrogate modeling plays a pivotal role in reducing computational costs while preserving predictive accuracy \cite{humfeldMachineLearningFramework2021}. The integration of machine learning (ML) techniques as surrogate models within Bayesian inference frameworks has enabled efficient (inverse) uncertainty quantification across diverse domains, including materials engineering, where such models support mechanical property inference, process parameter optimization, and quality prediction under uncertainty \cite{lepickaInitialEvaluationPerformance2019, malleyPredictabilityMechanicalBehavior2022, RAILLON2018818}, as well as in chemical process industries, where they facilitate thermodynamic model calibration \cite{PAULSON201974}.
	
	Among the widely used approaches that integrate the computational efficiency of surrogate modeling with the robustness of Bayesian inference is Approximate Bayesian Computation (ABC) \cite{Tavare1997, Beaumont2002, Marjoram2003}, a likelihood-free method that has gained considerable attention for model calibration under uncertainty, offering particular advantages when dealing with complex process where likelihood functions are intractable or computationally prohibitive \cite{Ratmann2009}. ML models, particularly decision tree-based models \cite{chenXGBoostScalableTree2016}, often exhibit intractable likelihoods due to their complex non-linear architectures. ABC algorithms address this limitation by utilizing summary statistics to approximate posterior distributions without requiring explicit likelihood evaluation. 
	
	The limited adoption of Approximate Bayesian Computation (ABC) for industrial-scale inverse uncertainty quantification highlights the need for robust and computationally efficient implementations. This work addresses these challenges through several key contributions. First, we integrate XGBoost regressors within the ABC framework, establishing a systematic approach for inferring credible intervals of critical process parameters in complex industrial systems. Second, we introduce an efficient sampling strategy for mixed-type data based on kernel-weighted distance metrics, enabling more effective exploration of high-dimensional parameter spaces while preserving computational tractability.
	
	A key innovation lies in the integration of this framework with \textit{Doc2Vec} embeddings for categorical variable representation \cite{ref:Doc2Vec, lau2016empiricalevaluationdoc2vecpractical}. In our case study, these embeddings guide the analysis by enabling effective handling of complex geometric descriptions through dense vectors in continuous vector spaces, facilitating similarity assessment and clustering based on embeddings \cite{8718215}. Furthermore, the framework incorporates a comprehensive validation methodology specifically designed for industrial applications, ensuring robust performance assessment under realistic operating conditions using actual production data from CVD processes.
	
	The methodology presented here advances both the theoretical understanding of inverse UQ in industrial systems and the practical application of ML and NLP techniques to process optimization. By utilizing actual production data containing inherent noise and uncertainties, in addition to those arising from the predictive model itself, the framework demonstrates how these methods can address traditional challenges in industrial process modeling while maintaining the interpretability and uncertainty quantification essential for engineering decision-making.

\section{Methods}\label{secc:Methods}
		
Uncertainty quantification merges two complementary paradigms: forward UQ propagates input uncertainties through a model to assess output variability, while inverse UQ addresses the inverse problem of estimating parameter distributions from observed data (see \cref{fig:Model}). In this section, we provide a description of the methods and algorithms used in the implementation of a data-driven Bayesian aproach for inverse uncertainty quantification. This approach combines prior knowledge about parameters with observational evidence through a likelihood function associated with a predictive (ML) model to obtain posterior distributions via Bayes' theorem. However, when we implement ML models in industrial-scale aplications, likelihood functions often become intractable, which makes it necessary for these models to adopt likelihood-free approaches, such as the ABC algorithm.

	\subsection{Problem Formulation}
		
\begin{figure}[!ht]
		\centering
		\includegraphics[width=1\linewidth]{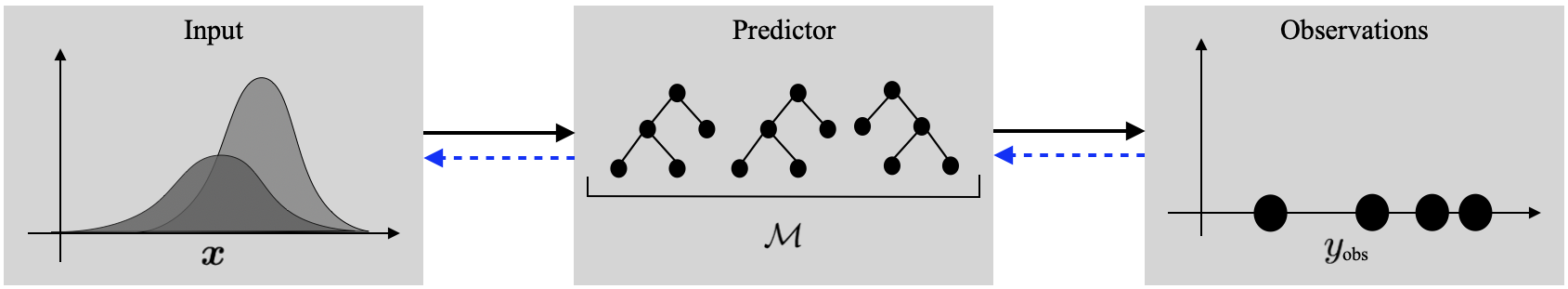}
		\caption{Forward uncertainty quantification (black solid line) . Inverse uncertainty quantification (blue dashed line).}
		\label{fig:Model}
	\end{figure}

	Let $\bm{x} \in \mathbb{R}^N$ denote the vector of input parameters, composed of continuous, integer, and binary components, where the binary variables take values in $\{0, 1\}$. Let $y_\text{obs} \in \mathbb{R}$ be an observable output variable. Let $\mathcal{M}: \mathbb{R}^N \to \mathbb{R}$ be a ML-based surrogate model (predictor, see \cref{fig:Model}) with intractable likelihood, such that
	\begin{align}\label{equation}
		y_\text{obs} = \mathcal{M}(\bm{x})+\varepsilon, \qquad \varepsilon \sim \pi_\varepsilon,
	\end{align} where $\varepsilon>0$ denotes the prediction error (residuals), \textit{i.e.}, $\mathcal{M}$ approximates the map between $\bm{x}$ and $y_\text{obs}$ (noisy model).
		
	\subsection{Predictor  using XGBoost}\label{Subsec:XGB}
		
	XGBoost (Extreme Gradient Boosting) is a decision tree-based method \citep{hastieEnsembleLearning2009, jamesTreeBasedMethods2021} that allows both classification and regression. This method builds shallow trees sequentially, with each tree fitted using error residuals from the previous model. Moreover, XGBoost can handle mixed input parameter types (numerical and categorical). This approach offers several advantages for the implementation of an inverse problem such as 
		
	\begin{itemize}
		\item \emph{Mixed Parameter Handling}: XGBoost naturally accommodates numerical and categorical input types without requiring extensive preprocessing.
		\item \emph{Computational Efficiency}: The algorithm provides speed improvements while maintaining prediction accuracy.
		\item \emph{Robustness}: The ensemble approach reduces overfitting and improves generalization.
	\end{itemize}
		
	However, this method also presents some limitations: as most of the ML models, XGboost is non-differentiable and likelihood-free, necessitating alternative approaches for uncertainty quantification and parameter estimation (refer to \cref{Subsec:ABC}). For additional details  on XGBoost, the reader is referred to \citet{chenXGBoostScalableTree2016}.
		
	We employ as our primary predictive model a XGBOOST regressor that in the following is denoted by $\mathcal{M}_\text{XGB}$.

	\subsubsection{XGBoost Feature Importance}
		
	Feature importance analysis via XGBoost built-in importance metrics that can be used to reduce the parametric space and identify critical parameters. More precisely, after constructing the boosted trees, we extract importance scores using the \texttt{total\_gain} metric \citep{UGUZ20111024}, which represents the total contribution of each feature across all trees. A higher \texttt{total\_gain} value indicates greater significance in generating predictions.

	\subsection{Information-Based Model Selection}\label{Subsec:modelselection}
		
	To identify the optimal probabilistic distribution that follows a random parameter, we implement an information-theoretic model selection approach \citep{GIOVANIS2022103293}. Given a set of realizations $\bm{x}$ of an input parameter $X$, we fit candidate models $\pi(\bm{x} \mid \theta)$ and select the best model using information criteria.
		
	\begin{itemize} 
		\item \textit{Model Selection Criteria:} We consider the \textit{Akaike Information Criterion (AIC)} defined as follows: $$\text{AIC} = -2 \log \mathcal{L}(\hat{\theta}) + 2k,$$
		where $\mathcal{L}(\hat{\theta})$ is the maximum likelihood of the fitted model, and $k$ is the number of model hyperparameters.
	\end{itemize}
		
	We call the best model to the one that minimizes the AIC.
		
	The probability that model $\pi^*(\bm{x} \mid \theta)$ best fits the data $\bm{x}$ is defined as:
	\begin{equation}
		P\big(\pi^*(\bm{x} \mid \theta)\big) \propto \exp\left(-\frac{\text{AIC}}{2}\right).
	\end{equation}
	The selected best model serves as the prior distribution $\pi$ in our Bayesian framework. Thus, we need to infer the best hyperparameters of the probabilistic model $\theta$.
	\begin{figure}[!ht]
		\centering
		\includegraphics[width=1\linewidth]{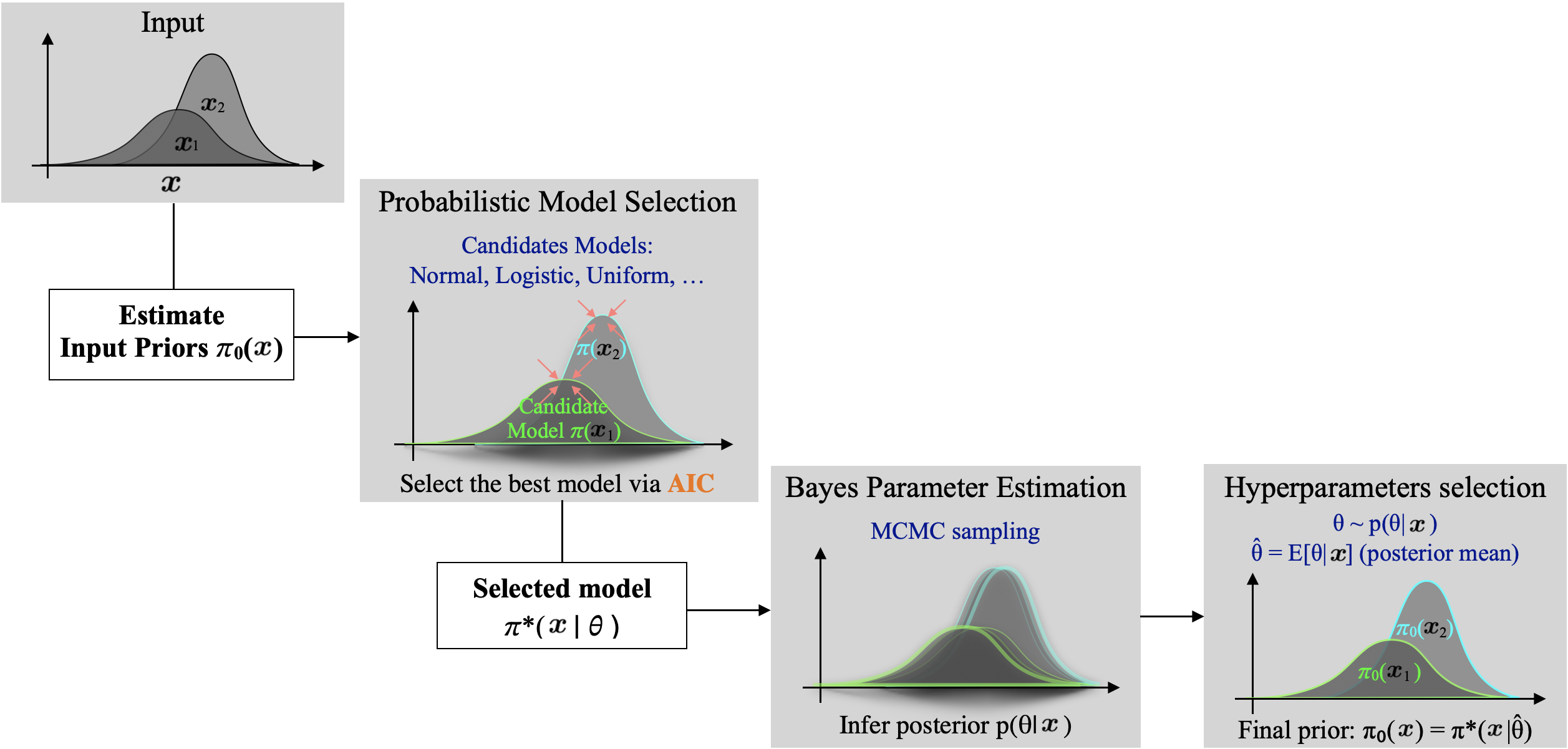}
		\caption{Bayesian workflow for data-driven prior estimation. Parameter data ($\bm{x}$) are fitted to candidate probability distributions, with the optimal model selected via AIC. Bayesian Parameter Estimation using MCMC sampling infers hyperparameter posteriors $p(\theta|\bm{x})$, and posterior mean values $\hat{\theta}$ define the final prior distributions $\pi_0(\bm{x}|\hat{\theta})$ for ABC inference.}
		\label{fig:Modelselection}
	\end{figure}

	\subsection{Bayesian Parameter Estimation}\label{Subsec:parameterestimation}
	Given a set of realizations $\bm{x}$ of an input parameter $X$, a parameterized (statistical distribution) model $\pi(\bm{x} \mid \theta)$, identified in \cref{Subsec:modelselection}, and a prior probability density function for model hyperparameters $\pi(\theta)$, we employ Markov Chain Monte Carlo (MCMC) to sample from the posterior probabilistic density function of the model hyperparameters:
	\begin{equation}
		\pi(\bm{x} \mid \theta) = \frac{p(\theta\mid \bm{x}) \pi(\theta)}{\pi(\bm{x})}.
	\end{equation}
	This approach enables robust parameter estimation while accounting for parameter uncertainty.
		
	\subsection{Inverse Uncertainty Quantification}\label{Subsec:InvUQ}
		
	Upon performing inverse UQ, we formulate an inverse problem that consists of inferring unknown parameters $\bm{x}$ from noisy measurements $y_\text{obs}$. 
	The solution to Bayesian inference is characterized by the Bayes' formula: $$\pi(\bm{x}|y_\text{obs}) \propto \mathcal{L}(y_\text{obs}|\bm{x}) \pi(\bm{x}),$$
		
	where $\pi(\bm{x})$ denotes the prior distribution, $\mathcal{L}(y_\text{obs}|\bm{x})$ is the likelihood function, and $\pi(\bm{x}|y_\text{obs})$ represents the posterior distribution.
		
	However, in most industrial applications involving machine learning predictors, the likelihood function is intractable due to the non-differentiable and complex nature of the models. 

	\subsubsection{ABC algorithm with Summary Statistics}\label{Subsec:ABC}
		
	Approximate Bayesian Computation represents a class of likelihood-free inference methods \cite{Beaumont2002, Marjoram2003, Ratmann2009, Sisson2007, Tavare1997} that circumvent the need for explicit likelihood evaluation through simulation-based approximation. ABC methods aim to calculate an approximation of the true posterior distribution $\pi(\bm{x}|y_\text{obs})$ by generating synthetic data $y_\text{sim}$ using a forward model $\mathcal{M}_\text{XGB}$, parameter values $\bm{x}$ sampled from the prior distribution $\pi(\bm{x})$ and residuals sampled from $\pi_\varepsilon$, through formula \eqref{equation}.
		
	\begin{figure}[!ht]
		\centering
		\includegraphics[width=1\linewidth]{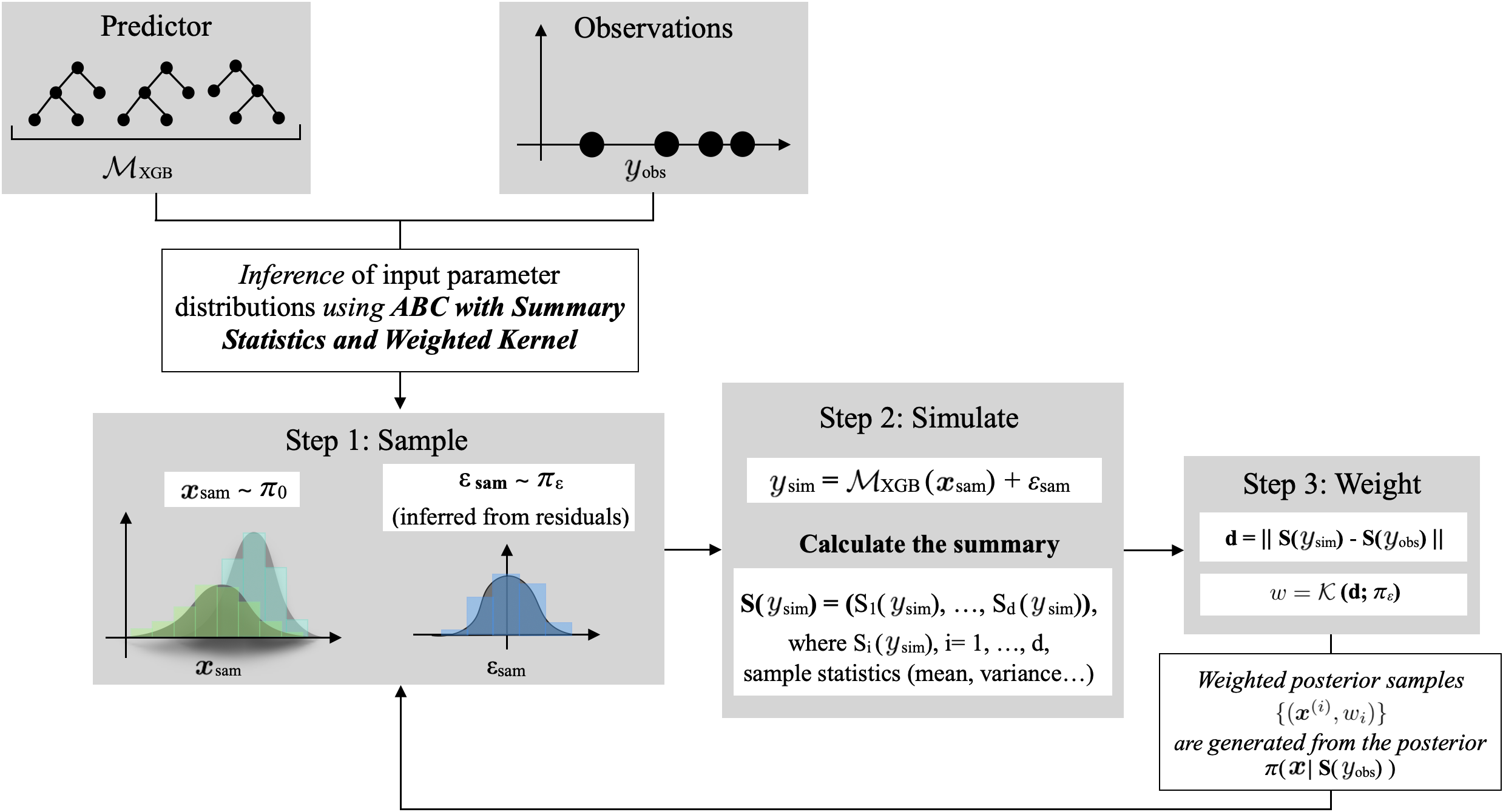}
		\caption{Weighted ABC inference framework. The surrogate model $\mathcal{M}_\text{XGB}$ and observations $y_\text{obs}$ enable parameter inference through: (Step 1) sampling from priors $\pi_0$ and error distribution $\pi_\varepsilon$, (Step 2) generating simulated data and computing summary statistics S($y_\text{sim}$), and (Step 3) assigning weights to all samples based on distance $d = \left\| S(y_\text{sim}) - S(y_\text{obs})\right\| $ using a kernel function $\mathcal{K}$ matching the characterized error distribution.}
		\label{fig:Alg}
	\end{figure}
		
	We implement a weighted ABC framework using kernel-weighted summary statistics. In contrast to rejection ABC, which discards all samples beyond a fixed tolerance threshold, the weighted approach assigns importance weights to all simulations based on their distance from observed data. This strategy improves computational efficiency by extracting information from the entire simulation ensemble. The weight function is defined as follows
	$$w_i = \mathcal{K}\left(d_i, \pi_\varepsilon\right), \qquad \text{with } d_i = d\big(S(y_\text{sim}^{(i)}), S(y_\text{obs})\big),$$
	where $d(\cdot,\cdot)$ is the Euclidean distance, $S(\cdot) = (S_1(\cdot), \ldots, S_d(\cdot))$ represents summary statistics (mean, standard deviation, median, and quartiles), and $\mathcal{K}$ is a kernel function parameterized by the error distribution $\pi_\varepsilon$. The error distribution characterizes the surrogate model residuals (prediction error), estimated from test data.
		
		
	The weighted posterior approximation is constructed as:
	$$\pi(\bm{x}|S(y_\text{obs})) \approx \sum_{i=1}^{N} \tilde{w}_i \cdot \delta_{x_i}(\bm{x}),$$
	where $N$ is the number of samples generated, $\tilde{w}_i = w_i / \sum_{j=1}^{N} w_j$ are normalized weights and $\delta_{x_i}$ is the Dirac delta function. The weighted sample $\displaystyle \{(x_i, w_i)\}_{i=1}^N$ approximates the posterior distribution $\pi(\bm{x} \mid S(y_\text{obs}))$, with samples exhibiting smaller distances receiving higher weights. Then, this weighted formulation ensures that suitable samples contribute more to the posterior approximation, enabling robust uncertainty quantification and sensitivity analysis. Furthermore, the effective sample size $\text{ESS} = 1/\sum_{i=1}^N w_i^2$ quantifies posterior concentration. 
		
	Finally, to ensure the posterior approximation with summary statistics $\pi(\bm{x}|S(y_\text{obs}))$ approaches the posterior $\pi(\bm{x}|y_\text{obs})$, the summary statistics must be sufficient \cite{Beaumont2002,10.121412STS406}. Sufficiency is achieved when $S(\cdot)$ captures all information in $y_\text{obs}$ relevant to $\bm{x}$, such that $\pi(\bm{x}|S(y)) = \pi(\bm{x}|y)$. In practice, we verify this by ensuring $S(\cdot)$ includes key distributional features (mean, variance, quantiles) of the observed data.

\section{Case Study: CVD Process Application}\label{sec:Process}
		
\begin{figure}[ht]
		\captionsetup[subfigure]{justification=centering}
		\centering
		\begin{subfigure}[b]{0.35\textwidth}
			\includegraphics[width=0.8\textwidth]{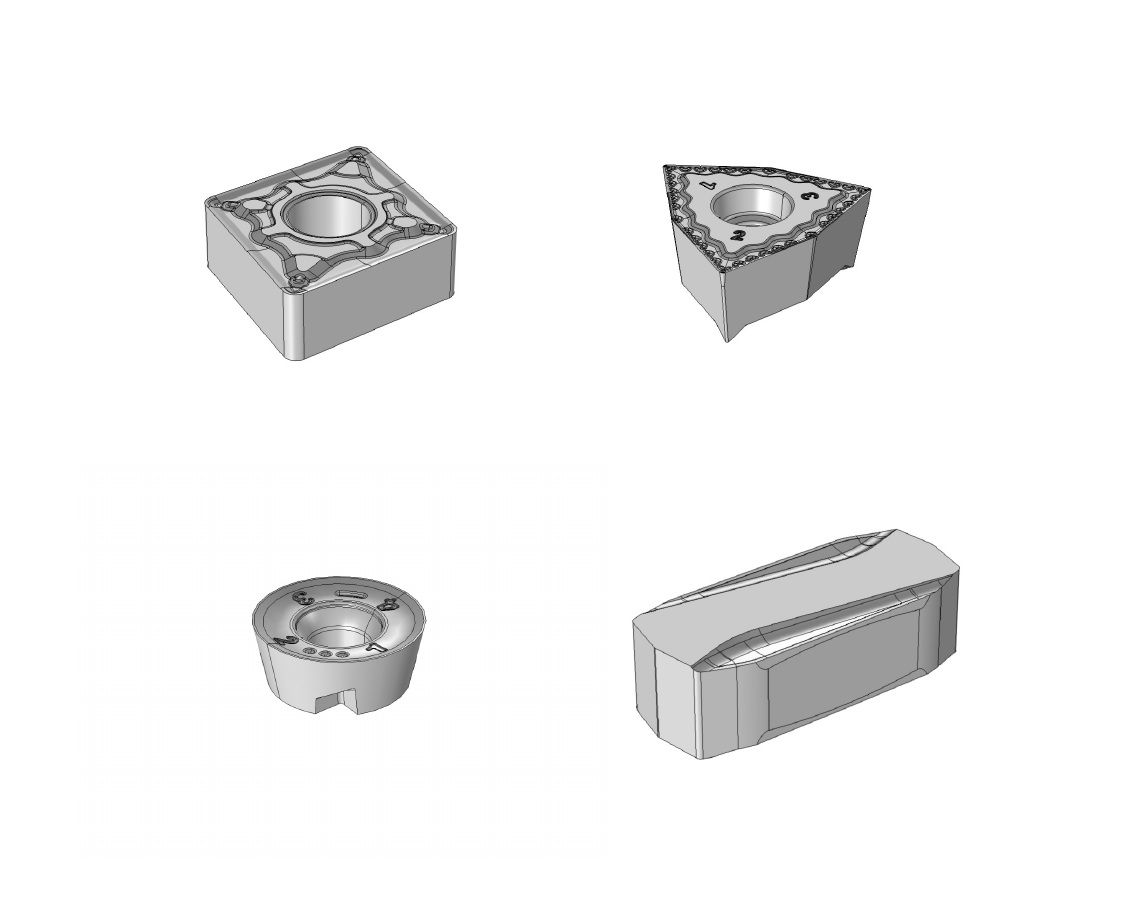}
			\caption{}
			\label{fig:inserts} 
		\end{subfigure}
		\hspace{2cm}
		\begin{subfigure}[b]{0.35\textwidth}
			\includegraphics[width=0.8\textwidth]{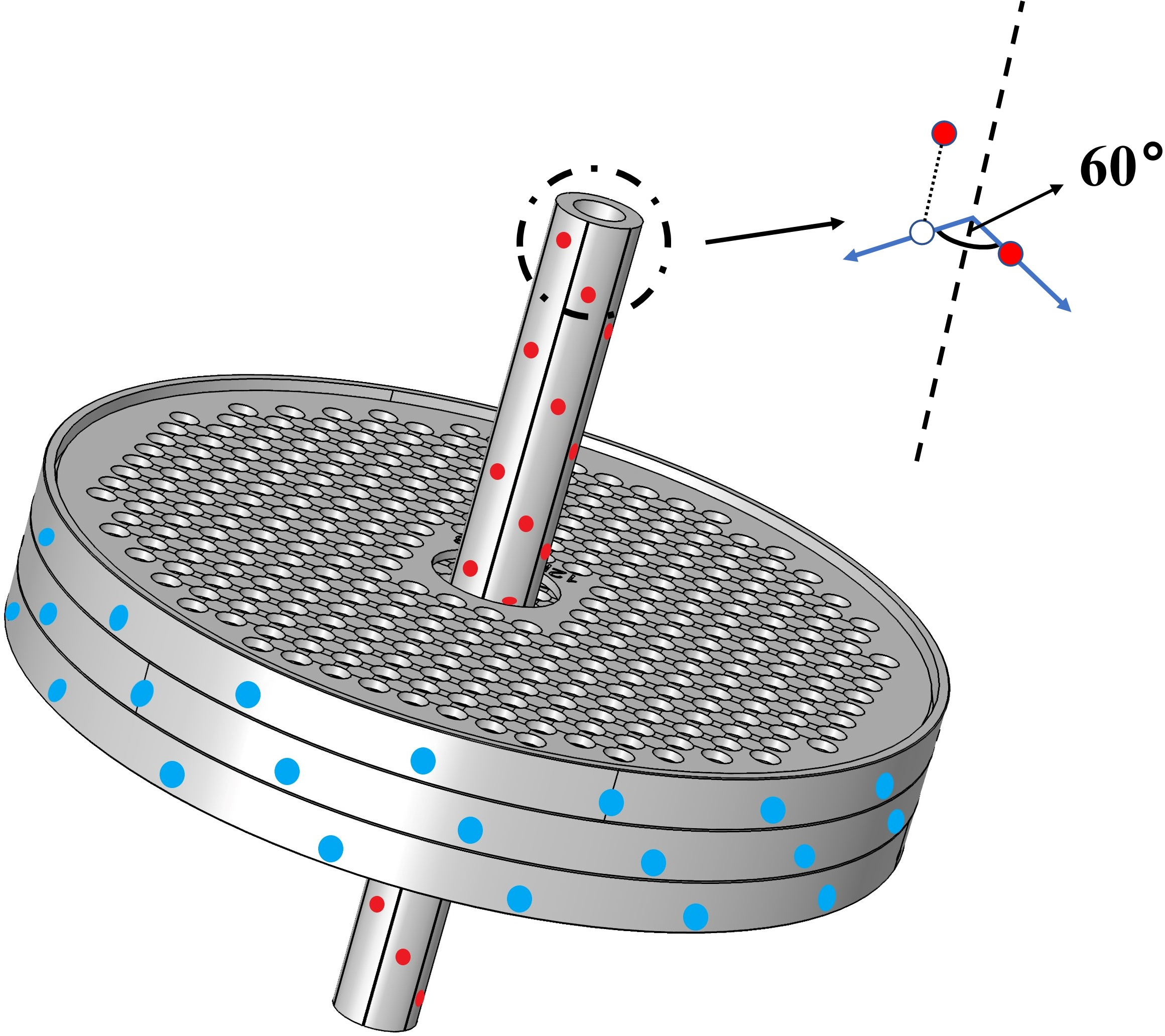}
			\caption{}
			\label{fig:3d-3disks}
		\end{subfigure}
		\caption{(a) Examples of the coated cutting tools. (b) A 3D representation of a 3-tray part of the reactor. The inlet perforations on the rotating inlet tube are shown in red. The outlet perforations for each tray are shown in blue.}
		\label{fig:geometry-explain}
	\end{figure}

	This study uses production data from a two-step coating process conducted in a commercial CVD reactor (Sucotec SCT600TH). The process involves sequential deposition of two layers on cemented carbide cutting inserts: (1) a 9 $\mu$m Ti(C,N) base layer (see \cref{fig:inserts}), followed by (2) an alumina (\alumina{}) layer deposited under an AlCl\textsubscript{3}–CO\textsubscript{2}–HCl–H\textsubscript{2}–H\textsubscript{2}S chemical system at $T$ = 1005°C and $p$ = 80 mbar \citep{hochauerCarbonDopedAAl2O32012}. This coating process and its variants have been extensively characterized in previous studies \cite{papavasileiou2024integrating, papavasileiou2022efficient, papavasileiouEquationbasedDatadrivenModeling2023}. 
		
	The CVD reactor consists of 40-50 perforated trays arranged in a vertical stack, each containing cutting inserts as illustrated in \cref{fig:3d-3disks}. Gas reactants are introduced through perforations in a rotating cylindrical tube positioned centrally within the tray stack. Each tray level has two diametrically opposite perforations with a 60\textdegree{} angular offset between consecutive levels. A critical aspect of this manufacturing environment is the variability in internal reactor geometry between production runs. The geometry of both the inserts and their supporting trays is modified according to specific production requirements. 
		
	\subsection{Data Collection}\label{sec:Data}
		
	The main objective of the coating process is achieving uniform thickness distribution, as coating uniformity directly correlates with product longevity and performance \citep{bar-henExperimentalStudyEffect2017,koronaki2014non}. Ideally, this uniformity would be consistent across all production runs, reactors, and sites. However, this is not always achieved. Then, establishing a systematic approach to evaluate factors affecting the uniformity of the coating thickness is essential. To this extent, the application of both equation-based methods \citep{papavasileiou2022efficient} and data-driven methods \citep{papavasileiouEquationbasedDatadrivenModeling2023, papavasileiou2024integrating} has been demonstrated in previous work, to which the interested reader is referred for further information on the process. In addition, previous works has implemented ML methods for predicting the coating thickness of the inserts based on the reactor setup \cite{papavasileiou2024integrating, KORONAKI2025109146}. 
		
\begin{figure}[!ht]
		\centering
		\includegraphics[width=.8\textwidth]{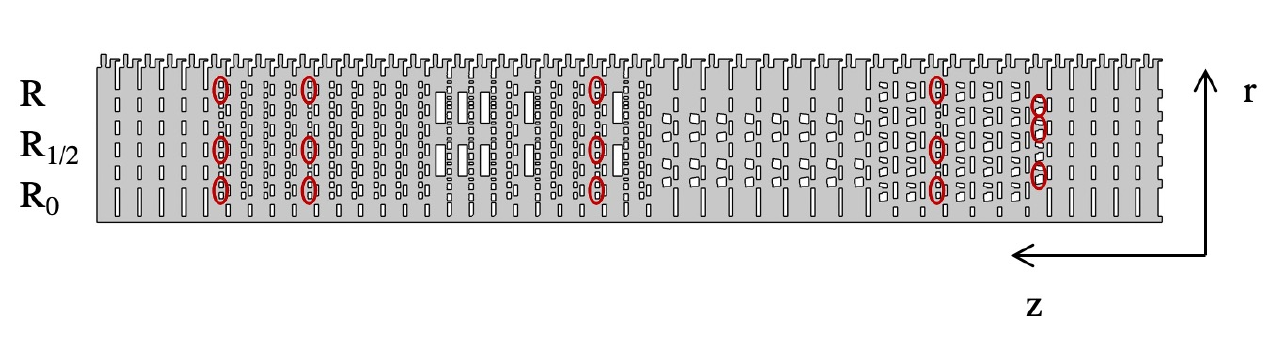}
		\caption{Positions with available \alumina{} thickness measurements from the production data for our test case. In general, across different production runs.}
		\label{fig:measurement-positions}
	\end{figure}

	At each production run, 15 thickness measurements are obtained ex-situ using the Calotest method \citep{lepickaInitialEvaluationPerformance2019}. These measurements are taken at predefined reactor positions, with additional measurements concentrated at the R position (closest to reactor outlet) due to production requirements (see \cref{fig:measurement-positions}). The measurement protocol ensures representative sampling across the reactor volume while accommodating practical production constraints.	
		
	In addition to coating thickness measurements, the dataset also contains several features regarding the set-up of the production run. These features are used as input parameters for the predictive ML model. An important feature is the production \enquote{recipe}, which encapsulates the steps taken and the process conditions during production. These specific details cannot be detailed here. In addition, the following set-up features are used: a) The number of inserts placed on each tray. b) The position of each tray within the reactor. c) The surface area of the inserts placed on each tray. d) The type of insert placed on each tray. Each type of insert has different geometrical characteristics. This last feature is represented using ISO designation codes \cite{ecatalog}, which are codes composed of eight alphanumeric characters encoding both numerical measurements and categorical descriptions of insert geometry.
		
	\begin{table}[ht]
		\centering
		\caption{Summary of reactor set-up features.}
		\begin{tabular}{ccc}
			\hline
			\textbf{Feature}                                                                           & \textbf{Type}       & \textbf{Pre-processing} \\
			\hline
			Number of inserts on tray                                                                  & Numerical (integer)& standardization        \\
			Tray position                                                                           & Numerical (integer)  & standardization        \\
			Surface area of inserts on tray                                                            & Numerical (float)   & standardization        \\
			\begin{tabular}[c]{@{}c@{}}Total surface area of inserts\\ inside the reactor\end{tabular} & Numerical (float)& standardization        \\
			Surface area standard deviation                                                            & Numerical (float)   &  standardization        \\
			\begin{tabular}[c]{@{}c@{}}Nominal “recipe” surface area \\ - actual surface area\end{tabular}    & Numerical (float)   &  standardization        \\
			Production \enquote{recipe}                                                                        & Categorical      & binary encoding        \\
			{\textit{Insert geometry}}                                                                            & Categorical     & binary encoding/embeddings        \\
			{\textit{Insert geometry} – tray above}                                                               & Categorical      & binary encoding/embeddings        \\
			{\textit{Insert geometry} – tray below}                                                               & Categorical      & binary encoding/embeddings        \\
			\hline
		\end{tabular}
		\label{table:feature-summary}
	\end{table}
		
	Additional features are engineered which include the total surface area and the standard deviation of the surface area of the to-be coated inserts. The information available for the neighboring trays, i.e. the trays above and below the tray of interest, are also used for the development of our predictive models. Subject matter expertise suggests that the difference between the nominal surface area indicated in the production \enquote{recipe} and the actual surface area of the inserts within the loaded reactor, is an important feature considered.
		
	In total, ten features, both numerical and categorical, are available for the development of the predictive model, as summarized in \cref{table:feature-summary}. The numerical features are standardized: centered (subtraction of the mean) and scaled (divided by the standard deviation), while the categorical variables are encoded using binary encoding or \textit{Doc2Vec} embeddings to maintain computational efficiency while preserving information content \citep{potdarComparativeStudyCategorical2017, KORONAKI2025109146}.

		\subsubsection{High-quality production}\label{sec:rank}
		
		Since the objective of this industrial production process is to ensure coating uniformity, it is necessary to define what constitutes a good production run. To accomplish this, we utilize the data collection gathered from each production run as described in the previous section. We define a production run as successful if the mean coating thickness produced meets production standards and its standard deviation is as low as possible.
		
		Thus, under this definition, for each production run, the respective coating thickness sample mean and standard deviation are calculated. Based on both quantities, the production batches are ranked. This ranking will allow us to use the conditions under which they were produced as a reference point for comparing and validating the results of subsequent analysis.
		\begin{figure}[!ht]
			\centering
			\includegraphics[width=.675\linewidth]{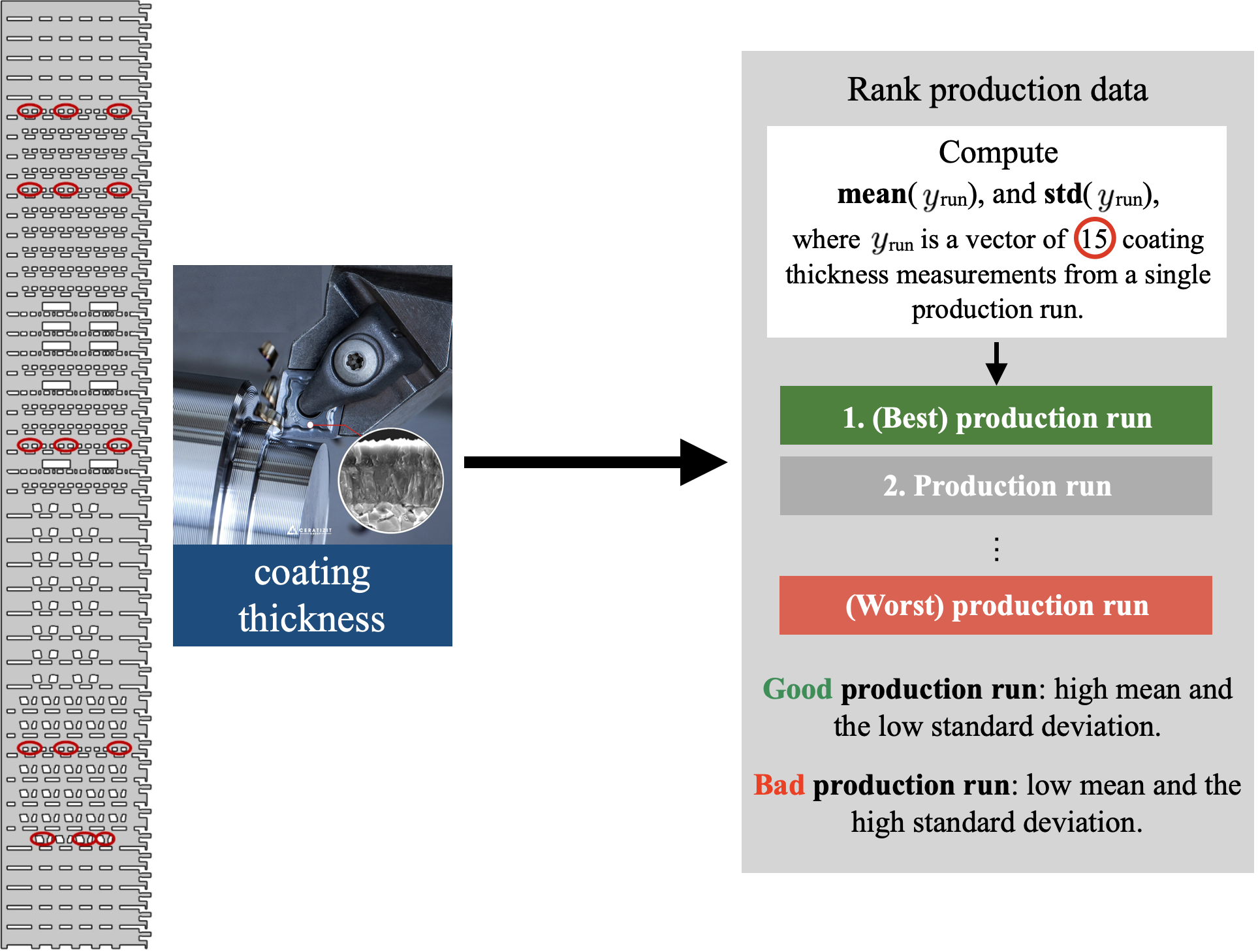}
			\caption{Production data is collected and ranked, with batch (production run) statistics: mean and standard deviation of coating thickness. Production runs are categorized based on two metrics: Run 1 (green) represents optimal quality with high mean coating thickness and low variability; intermediate runs (gray) show moderate performance; Run n (red) indicates deficient quality characterized by low mean thickness and high variability.}
			\label{fig:Last}
		\end{figure}

	\subsection{Proposed workflow}\label{Sec:workflow}
		
	For the implementation of Bayesian inference method, we need to manage two instances: learning and prediction. The list below presents the steps to follow:
	\begin{enumerate}
		\item \textit{Define a surrogate model}, base on the XGBoost regresor, that maps input parameters (related to the reactor set-up) to a output variable (e.g. insert thickness).
		\item \textit{Feature Importance analysis}, base-on the surrogate model using \texttt{total\_gain} metrics.  Select top-$q$ most important parameters for subsequent analysis. 
		\item \textit{Determine prior distribution}, applying information-based model selection to identify optimal probability distributions for each important input parameter, and estimate their corresponding hyperparameters using Bayesian parameter estimation with MCMC.
		\item \textit{Infer the parameter posteriors distributions}, implementing ABC algorithm with a fix tolerance thresholds and summary statistics to generate samples from approximate posterior distributions.
		\item \textit{Validation and quantify uncertainties}, propagating input uncertainties through the forward model to the output variable and validate using quality definitions.
	\end{enumerate}

	\section{Results}\label{Results}
	
	It is worth noting that we are working with a large dataset that collects actual production data from the coating process described in \cref{sec:Process}. After processing categorical parameters, this dataset contains both continuous and discrete (integer and binary) parameters related to the reactor setup and the geometry of the cutting tools. Additionally, for each combination of input parameters, we have measurements of the coating thickness produced under that configuration and position. Therefore, we construct a surrogate model that maps combinations of input parameters to their corresponding thickness measurements. Then, we use Bayesian parameter estimation to infer posterior distributions of the input parameters given the observed data. Thus, taking into account the available observations, we are able to infer meaningful probabilistic information to calibrate the parametric space of the input parameters.

	\subsection{Data preprocessing}\label{sec:Res_Preprocessing}
	
	The dataset contains two main categorical features: production \enquote{recipe} and \enquote{insert geometry} distributed across three different trays. For the \enquote{insert geometry} features, each ISO designation code is decomposed into numerical features and textual descriptions to be use in the XGBoost predictor model training. The textual descriptions, comprising twelve different insert shape categories across tray positions, are processed using either binary encoding or \textit{Doc2Vec} embeddings. This allows us to assess the predictor performance under both representation schemes. The production \enquote{recipe} parameter is processed only through binary encoding and remains independent of insert geometry parameters, as reactors contain various insert shapes simultaneously. This configuration makes embedding-based representation particularly suitable for capturing geometric shape relationships. 
	\begin{figure}[!ht]
		\centering
		\includegraphics[width=.6\linewidth]{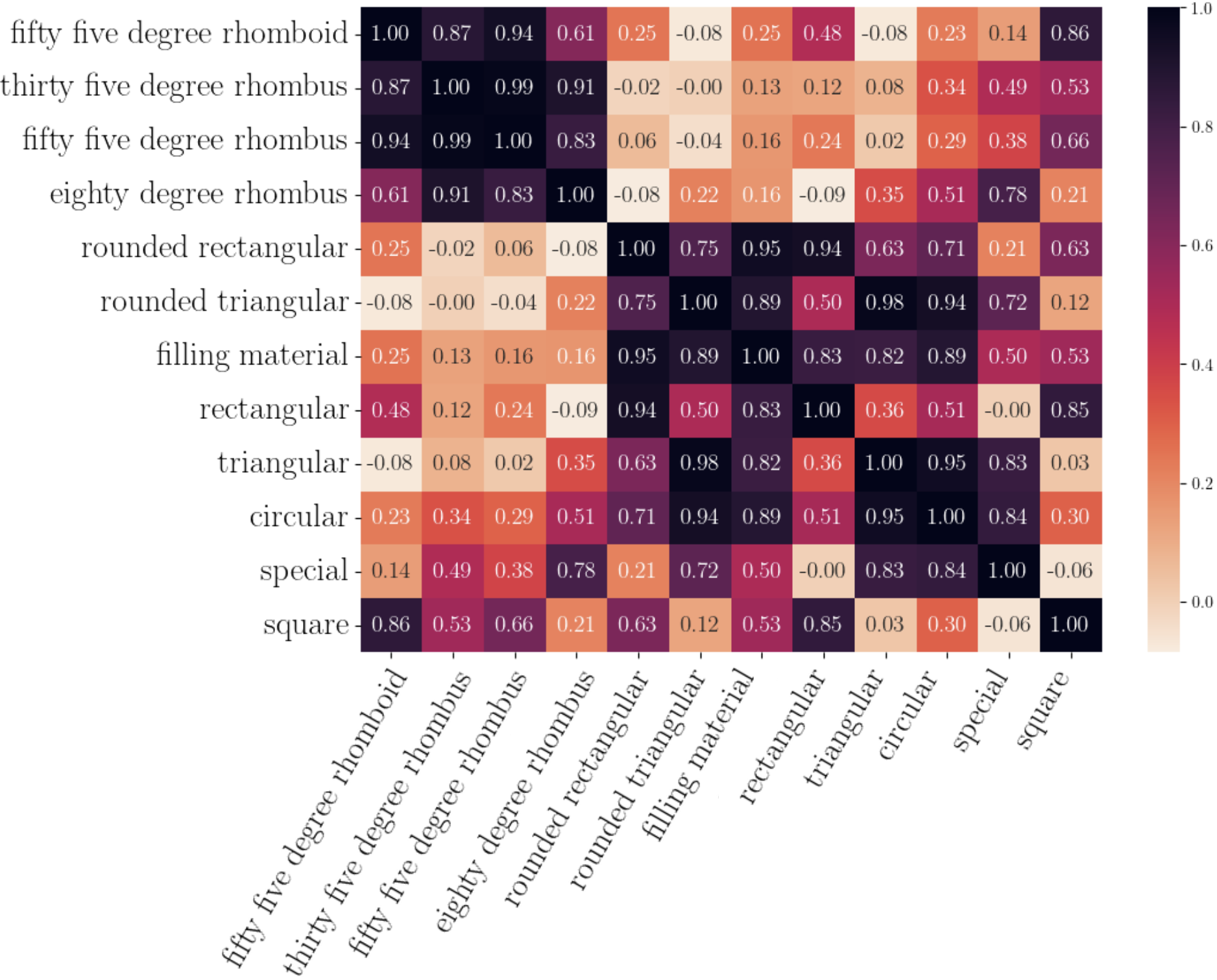}
		\caption{Heatmap showing the cosine similarity values calculated from the dense vectors obtained using the \textit{Doc2Vec} model. These values range from -1 to 1, where values closer to 0 are represented in lighter shades, and values closer to -1 or 1 are colored using darker shades.}
		\label{fig:cosine}
	\end{figure}
	
	To verify whether the embeddings capture the semantic and contextual relationships among different shape descriptions, we computed the cosine similarity matrix across all categories (see \cref{fig:cosine}). 
	
	The similarity scores demonstrate that the embeddings effectively represent shape relationships, with geometrically similar shapes exhibiting higher cosine similarity values. For instance, rhomboid-type shapes cluster together in the upper-left quadrant of the similarity matrix, confirming that the embedding space preserves meaningful geometric differences.

	\subsection{Surrogate Model Performance}\label{Subsec:SurrogateModel}
	
	A XGBoost regressor serves as the surrogate predictor, mapping process setup parameters into coating thickness, as to it has proven high performance for this task \cite{papavasileiou2024integrating, KORONAKI2025109146, papavasileiouEquationbasedDatadrivenModeling2023}. Dense vectors obtained by encoding/embedding categorical variables are included as numerical features in the XGBoost training. Model training employed $k$-fold cross-validation ($k = 10$) with performance evaluation through R$^2$, MSE, and MAE metrics.
	
\begin{figure}[!ht]
		\centering
		\begin{subfigure}[b]{0.35\textwidth}
			\includegraphics[width=1\textwidth]{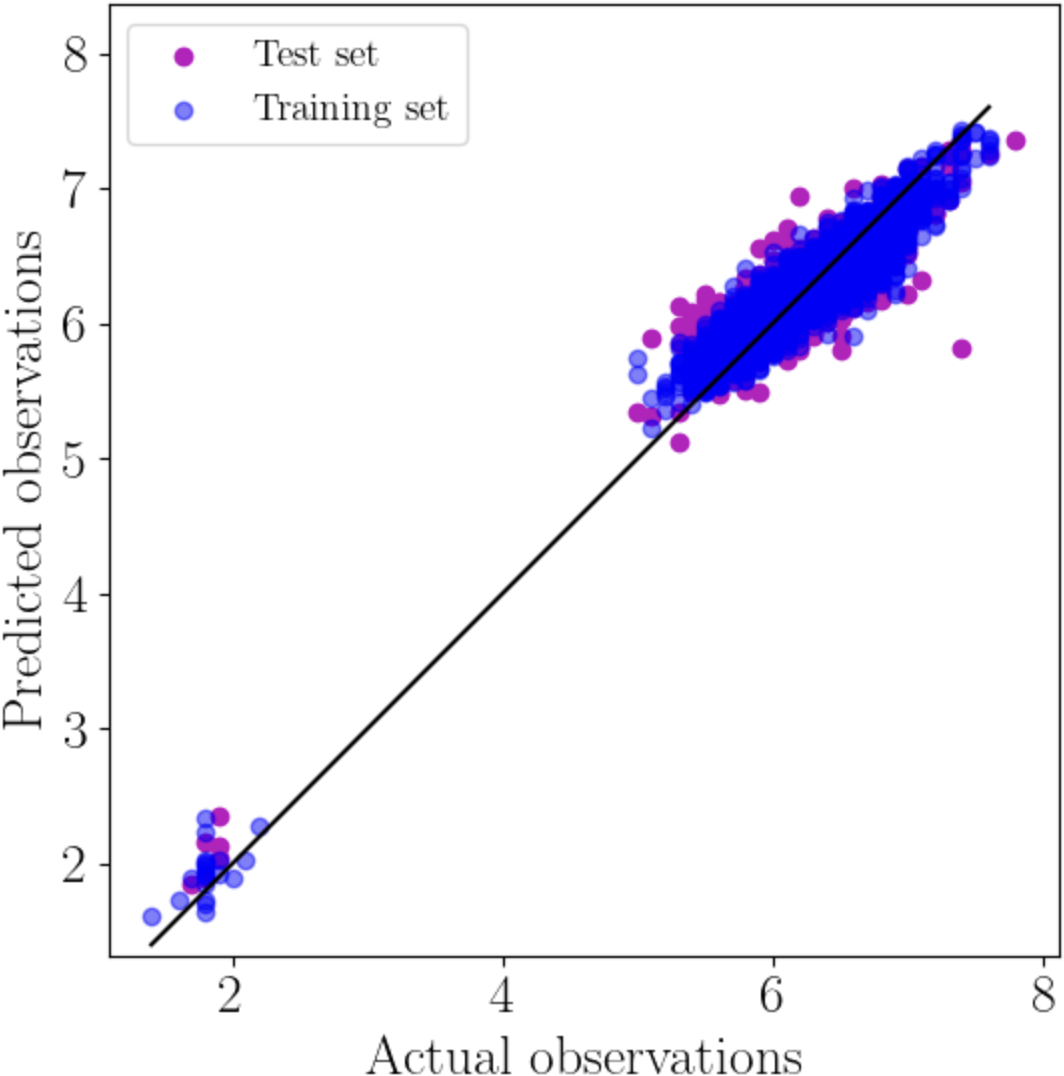}
			\caption{}
			\label{fig:inserts1} 
		\end{subfigure}
		\hspace{2cm}
		\begin{subfigure}[b]{0.35\textwidth}
			\includegraphics[width=1\textwidth]{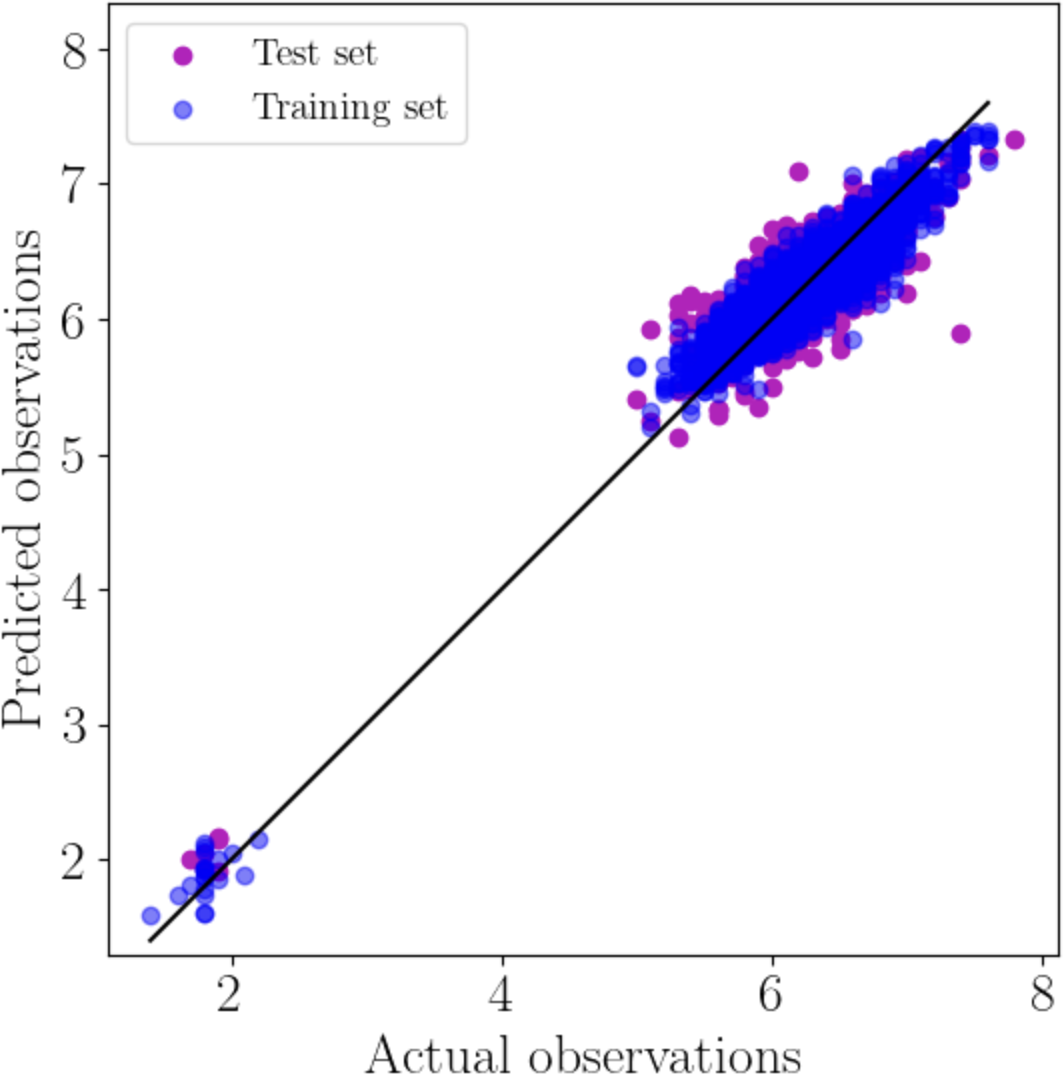}
			\caption{}
			\label{fig:inserts2} 
		\end{subfigure}
		\caption{Surrogate model performance comparison. (a) Model with binary encoding: R$^2$ = 0.777, MAE = 0.179. (b) Model with \textit{Doc2Vec} embedding: R$^2$ = 0.766, MAE = 0.182. The scatter plot compares predicted observations against actual observations for both training set (blue points) and test set (violet points). The diagonal line represents perfect prediction agreement.}
		\label{fig:xgbb}
	\end{figure}

	We compared two approaches for representing categorical variables related to the inserts geometries: binary encoding and \textit{Doc2Vec} embeddings derived from short textual descriptions. The surrogate model with binary encoding achieved R$^2$ = 0.777, MAE = 0.179, and MSE = 0.052 (see \cref{fig:inserts1}), demonstrating robust predictive capability. The model using \textit{Doc2Vec} embeddings yielded also high performance with R$^2$ = 0.766, MAE = 0.182, and MSE = 0.055 (see \cref{fig:inserts2}). In this case, the performance of this model is similar to that obtained using binary encoding; however, the true advantage of the continuous embedding representation lies in its ability to enhance surrogate prediction while enabling meaningful shape-based clustering for stratified ABC inference (see \cref{Subsec:Geometry}). Both models enable efficient likelihood-free inference, but \textit{Doc2Vec} provides high fidelity for capturing complex geometry relationships.
	
	Furthermore, feature importance was assessed using XGBoost feature importance metrics to ensure robust parameter selection for ABC implementation.

	\subsection{Feature importance Analysis}\label{Subsec:FI}
	
	Feature importance analysis, combined with expert knowledge, identifies critical parameters influencing coating quality, thereby guiding the application of Bayesian inverse uncertainty quantification to the most relevant process variables.
	
\begin{figure}[!ht]
		\centering
		\begin{subfigure}[b]{0.4\textwidth}
			\includegraphics[width=1\textwidth]{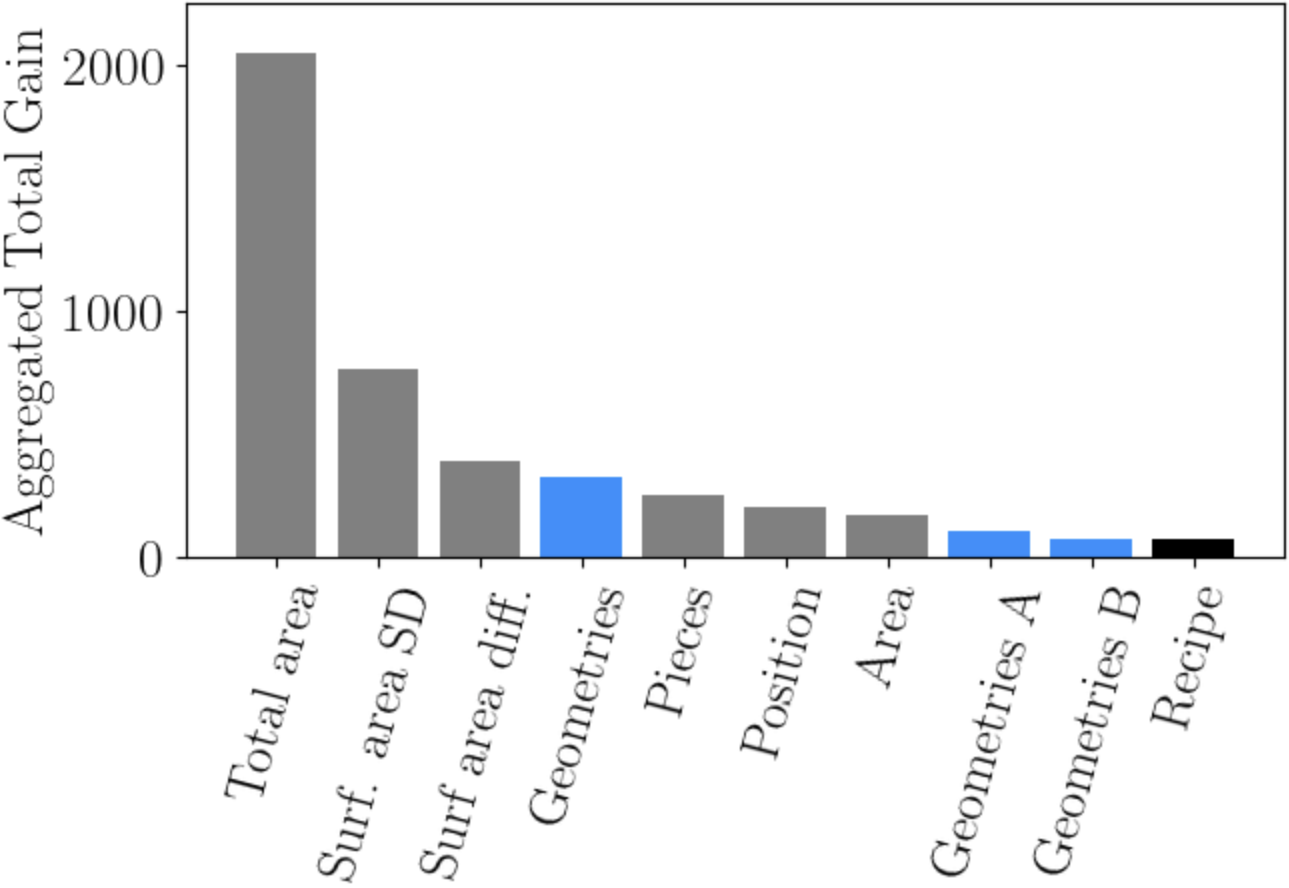}
			\caption{}\label{fig:IMPp1}
		\end{subfigure}
		\hspace{2cm}
		\begin{subfigure}[b]{0.4\textwidth}
			\includegraphics[width=1\textwidth]{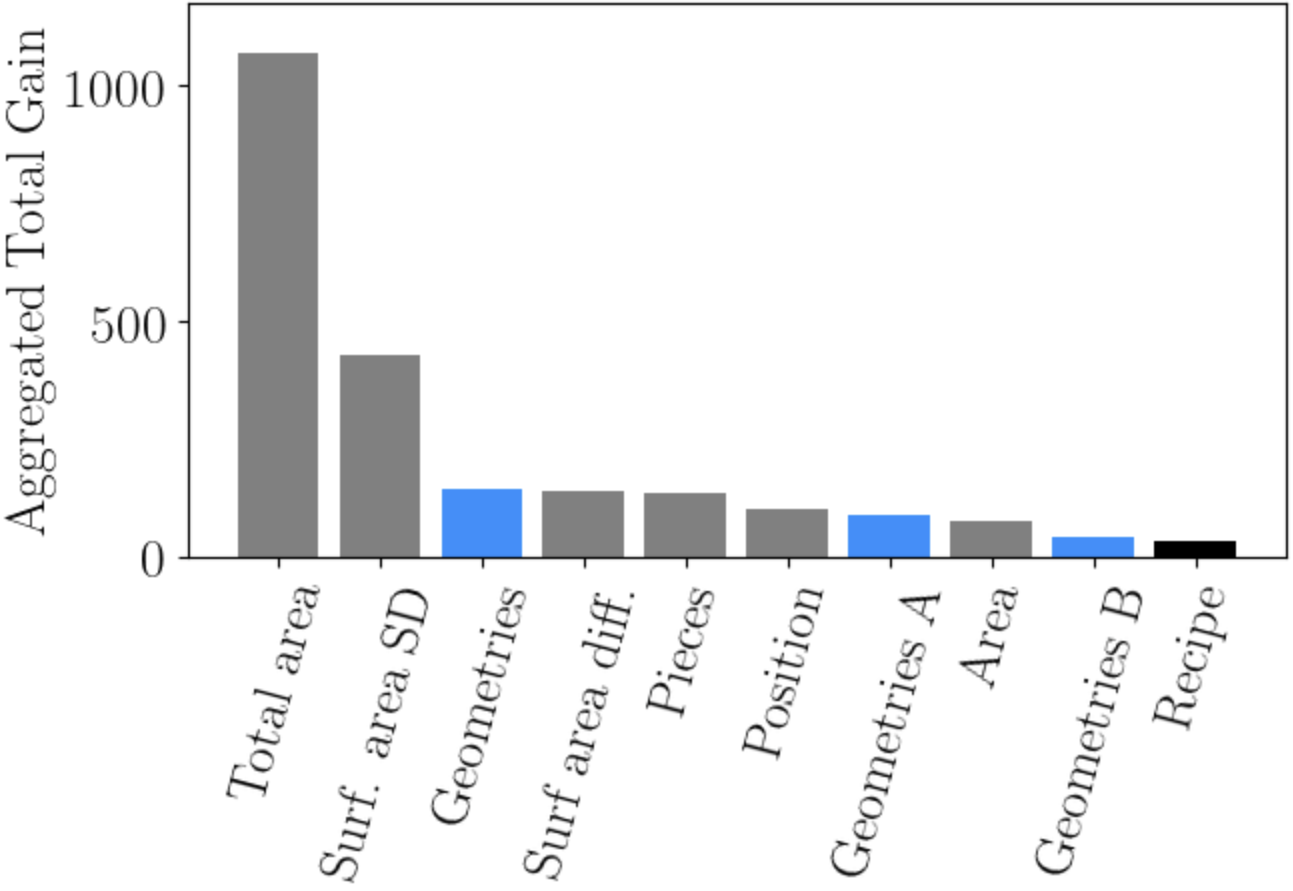}
			\caption{}\label{fig:IMPp2}
		\end{subfigure}
		\caption{Top ten input features with the highest importance scores measured by \texttt{total\_gain} for (a) Model with binary encoding and (b) Model with \textit{Doc2Vec} embedding. The gray bars represent numerical features, the black bars represent the categorical feature called "recipe", and the blue bars correspond to categorical variables indicating the geometries of the inserts located at the current position, above, and below.}
		\label{fig:IMPp}
	\end{figure}

	The XGBoost feature importance scores, based on \texttt{total\_gain}, reveal the relative contribution of each parameter to coating thickness prediction. When we employ binary encoding for processing categorical features, numerical parameters ("Total area", "Surface area SD", "Surface area diff.", "Position", "Pieces", "Area") dominate the importance ranking, while categorical parameters associated with insert geometries and production \enquote{recipe} after encoding exhibited lower scores (see \cref{fig:IMPp1}).
	
	However, the implementation of \textit{Doc2Vec} embeddings for categorical feature encoding transformed this scenario (see \cref{fig:IMPp2}), with embedded categorical features emerging as significant contributors to model predictions (particularly, "Geometries" and "Geometries Above"). This transformation enabled the importance quantification of this categorical variables that subject matter expertise suggested are critical in the process but remained underrepresented in binary encoding schemes. 

	\subsection{Information-based model selection and parameter estimation}\label{sec:IBBPE}
	
\begin{figure}[!ht]
		\centering
		\begin{subfigure}[b]{0.85\textwidth}
			\includegraphics[width=1\textwidth]{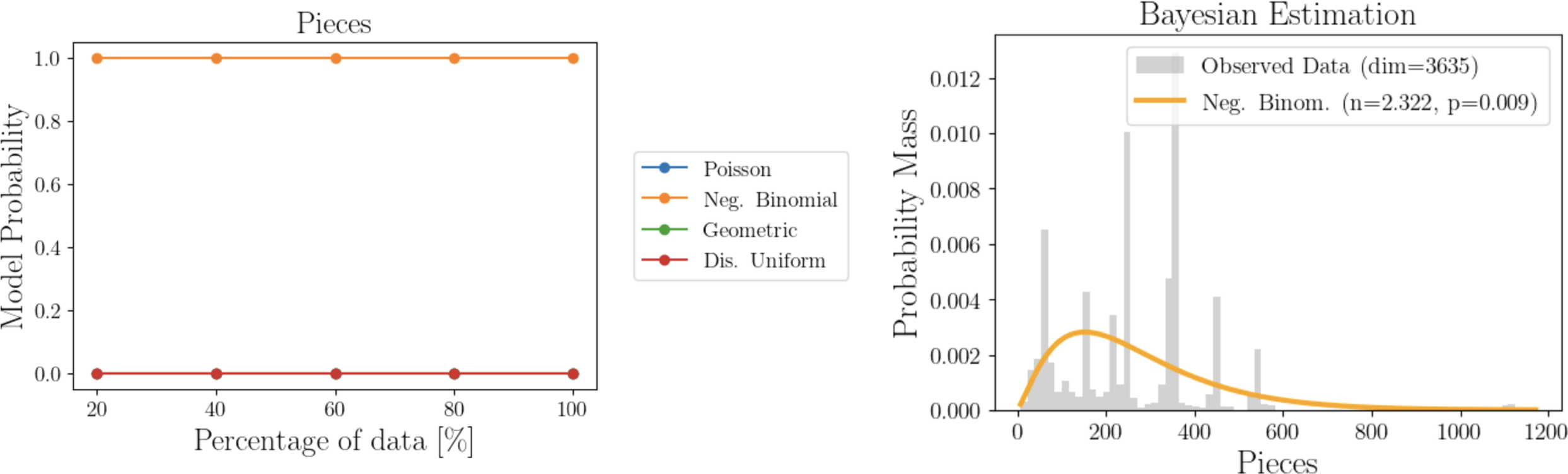}
			\caption{}
		\end{subfigure}
		\begin{subfigure}[b]{0.85\textwidth}
			\includegraphics[width=1\textwidth]{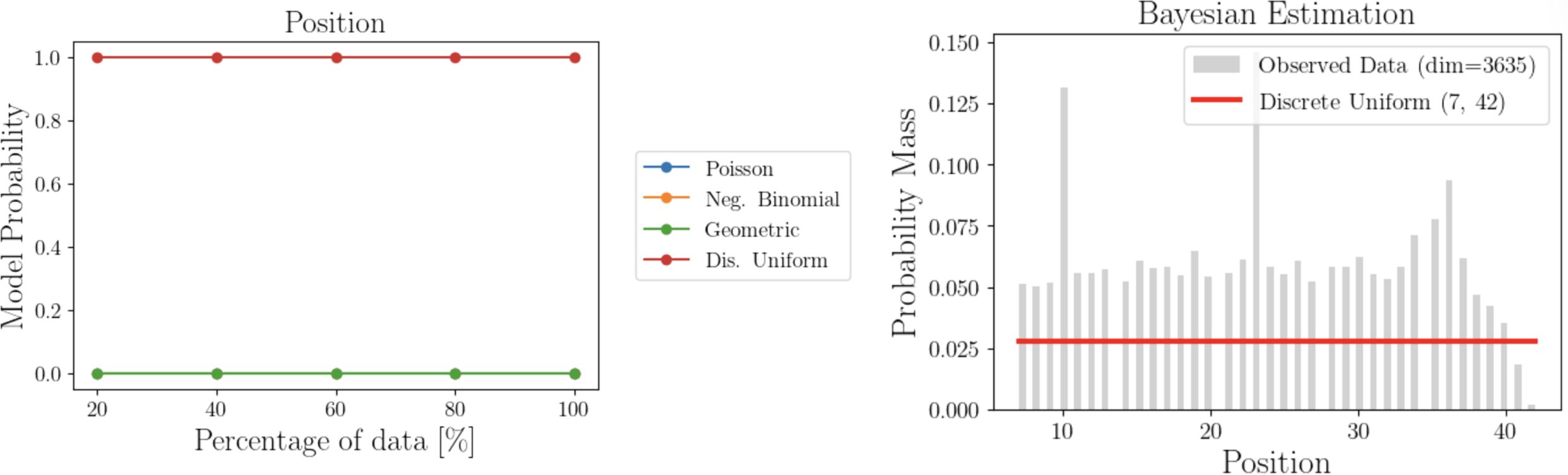}
			\caption{}
		\end{subfigure}
		\caption{Model selection and parameter estimation results for discrete setup parameters. Left: Model probabilities computed via information criteria showing clear preference for optimal statistical distributions. Right: Posterior distributions of best-fitting models (a) for "Pieces" (orange) and (b) for "Position" (red) with observed data histograms (gray bars), validating model-data agreement.}
		\label{probmodeldiscB}
	\end{figure}

\begin{figure}[!ht]
		\centering
		\begin{subfigure}[b]{0.85\textwidth}
			\includegraphics[width=1\textwidth]{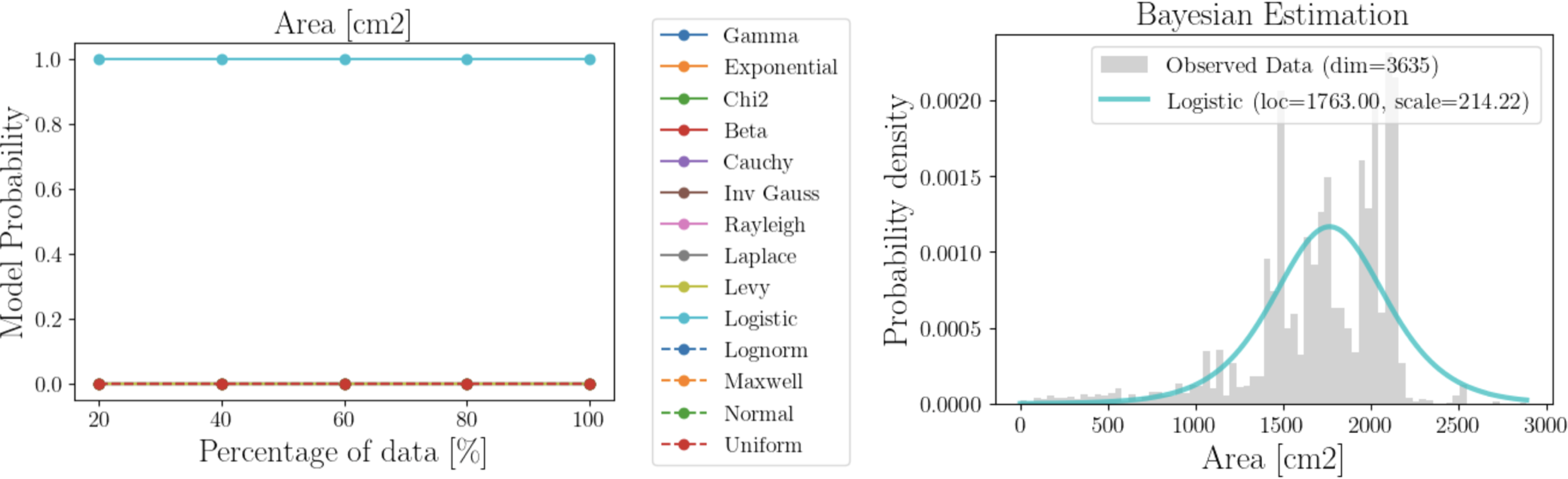}
			\caption{}
		\end{subfigure}
		\begin{subfigure}[b]{0.85\textwidth}
			\includegraphics[width=1\textwidth]{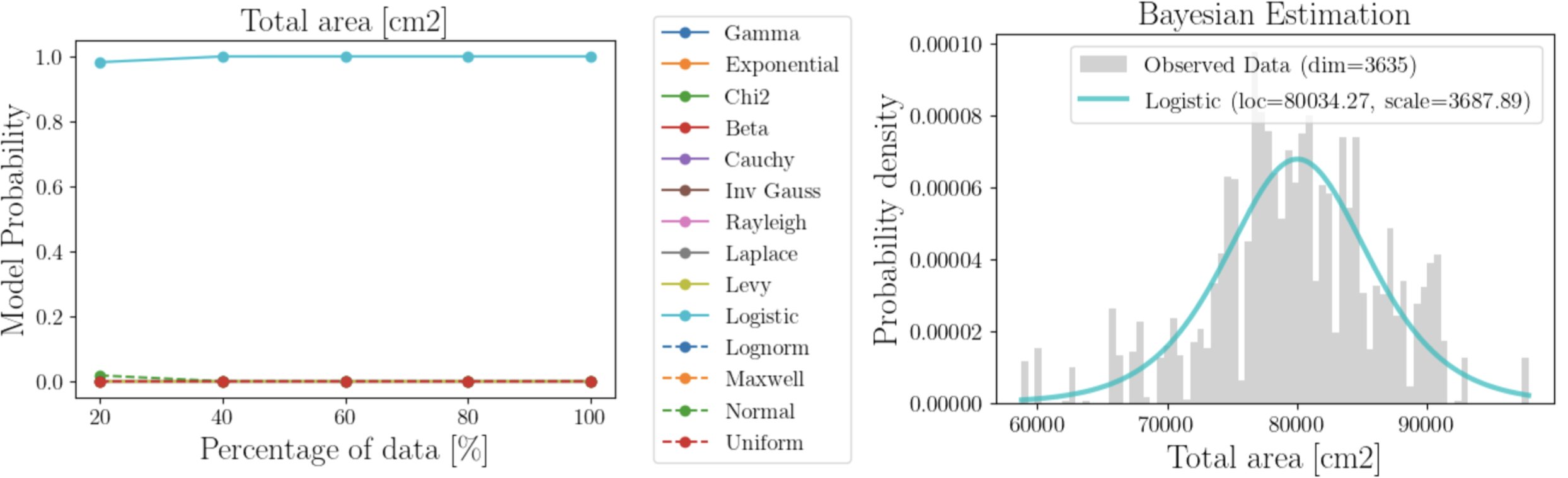}
			\caption{}
		\end{subfigure}
		\begin{subfigure}[b]{0.85\textwidth}
			\includegraphics[width=1\textwidth]{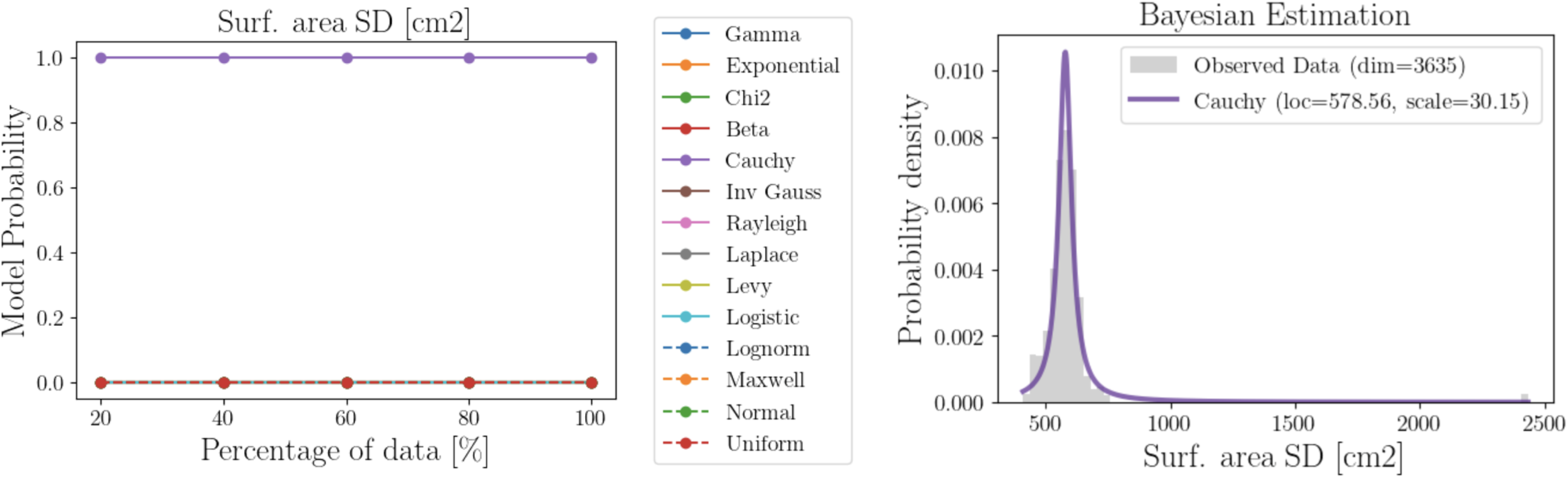}
			\caption{}
		\end{subfigure}
		\begin{subfigure}[b]{0.85\textwidth}
			\includegraphics[width=1\textwidth]{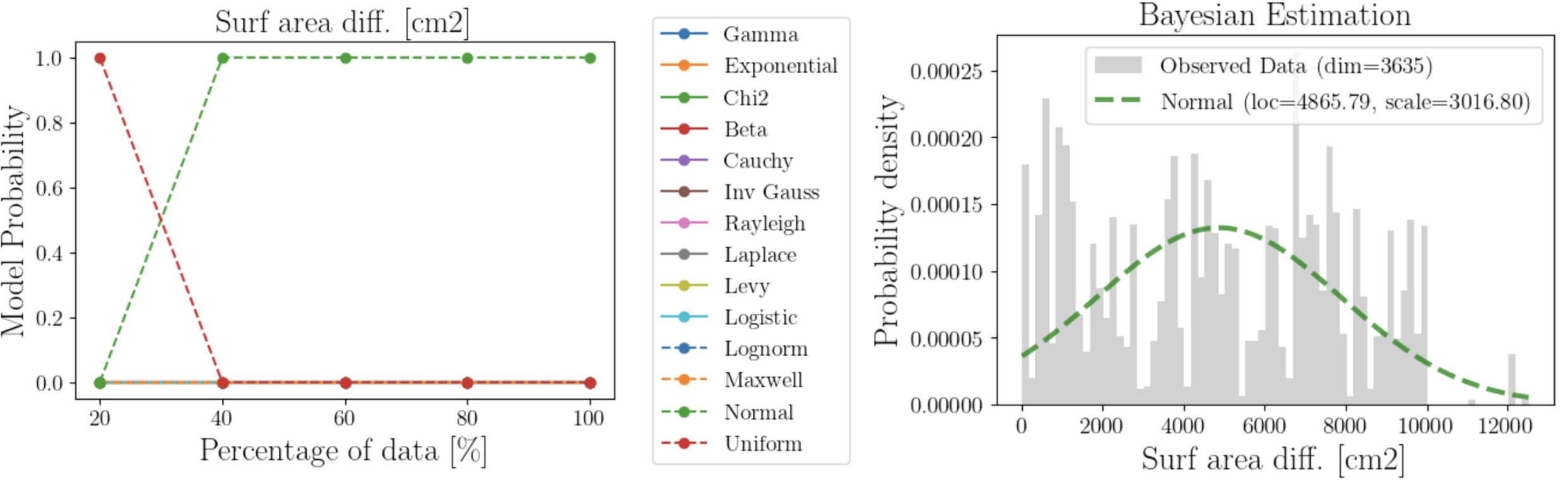}
			\caption{}
		\end{subfigure}
		\caption{Model selection and parameter estimation results for continuous setup parameters. Left: Model probabilities computed via information criteria showing clear preference for optimal statistical distributions. Right: Posterior distributions of best-fitting models (a) for "Area" (light blue), (b) for "Total area" (light blue), (c) for "Surface area standard deviation" (purple) and (d) for "Surface area difference" (green) with observed data histograms (gray bars), validating model-data agreement.}
		\label{probmodelcontB}
	\end{figure}

	Model selection was performed using information-theoretic criteria to evaluate probabilistic models across different data partitions. The selected model achieved consistently higher marginal likelihood across all partitions. The analysis demonstrates robust model selection for all parameters, with the chosen model having the highest probability among candidate models (see Figs. \ref{probmodeldiscB} and \ref{probmodelcontB}), indicating a clear preference for the optimal configuration. 
	
	\begin{table}[!ht]
		\centering
		\caption{Prior distributions for input parameters estimated via Bayesian Parameter Estimation. Distribution types were selected using AIC, and hyperparameters represent MCMC posterior means.}
		\begin{tabular}{ccc}
			\hline
			\textbf{Input parameters}                                                                           & \textbf{Probability Distribution}       & \textbf{Hyperparameters} \\
			\hline
			Number of inserts on tray                                                                  & Negative binomial & $n = 2.322,\ p = 0.009$\vspace{0.15 cm}       \\
			Tray position                                                                           & Discrete Uniform  & $a = 7,\ b = 42$      \\
			Surface area of inserts on tray                                                            & Logistic & $\mu = 1763.00,\ s = 214.22$\vspace{0.15 cm}        \\
			\begin{tabular}[c]{@{}c@{}}Total surface area of inserts\\ inside the reactor\end{tabular} & Logistic& $\mu = 80034.27,\ s = 3687.89$       \\
			Surface area standard deviation                                                            & Cauchy &  $x_0 = 578.56,\ \gamma = 30.15$\vspace{0.15 cm}        \\
			\begin{tabular}[c]{@{}c@{}}Surface area difference\end{tabular}    & Normal &  $\mu = 4865.79,\ \sigma = 3016.80$\vspace{0.15 cm}        \\
			\begin{minipage}{3.5cm}{\begin{center}Production \enquote{recipe} {\footnotesize (binary components)} \end{center}}\end{minipage}                                                                        & Binomial & \begin{minipage}{3.75cm}\begin{center}$p_0 =0.003,\ p_1=0.144,$\\ $p_2=0.426,\ p_3=0.875$\vspace{0.15 cm}\end{center}\end{minipage} \\
			\begin{minipage}{3cm}{\begin{center}\textit{Insert geometry} {\footnotesize (binary components)} \end{center}}\end{minipage}                                                                           & Binomial & \begin{minipage}{3.75cm}\begin{center}$p_0 =0.093,\ p_1=0.420,$\\ $p_2=0.355,\ p_3=0.615$\vspace{0.15 cm}\end{center}\end{minipage}\\
			\begin{minipage}{4cm}{\begin{center}\textit{Insert geometry} – tray above {\footnotesize (binary components)} \end{center}}\end{minipage}
			& Binomial & \begin{minipage}{3.75cm}\begin{center}$p_0 =0.093,\ p_1=0.352$\\ $p_2=0.577,\ p_3=0.424$\vspace{0.15 cm}\end{center}\end{minipage}\\
			\begin{minipage}{4cm}{\begin{center}\textit{Insert geometry} – tray below {\footnotesize (binary components)} \end{center}}\end{minipage}           & Binomial & \begin{minipage}{3.75cm}\begin{center}$p_0 =0.097,\ p_1=0.401,$\\ $p_2=0.542,\ p_3=0.723$\end{center}\end{minipage}\\
			\hline
		\end{tabular}
		\label{table:modeldistribution}
	\end{table}
	
	Parameter estimation across hyperparameter families generated probability distribution curves for setup parameters, with hyperparameter values summarized in \cref{table:modeldistribution}. Akaike information criteria consistently favored models that best fit the data, while the resulting distribution families demonstrated stability to hyperparameter choices. 
	
	This approach ensures robust parameter inference by accounting for model uncertainty and hyperparameter selection within a suitable probabilistic framework. The inferred models and hyperparameters are used as prior distributions in the subsequent analysis, as they provide a well-informed starting point for inference.
	
	A similar procedure is applied to the categorical parameters that were encoded using \textit{Doc2Vec} embeddings, and the residuals for both predictive models (with binary encoding and \textit{Doc2Vec} embeddings). The corresponding results can be found in Appendix \ref{Appendix1}, \cref{table:errorinf}.

	\subsubsection{Results Validation and Forward Prediction}\label{sec:Validation}
	
	Now, we analyze parameter inference results obtained through our Bayesian framework. The ABC implementation requires three components: a robust and accurate surrogate model (\cref{Subsec:SurrogateModel}), prior distributions for all continuous, discrete, and binary parameters inferred via information criteria (\cref{sec:IBBPE}, and the prediction error distribution, $\varepsilon \sim \text{Logistic}(\mu, s)$ with known hyperparameters (see Appendix \ref{Appendix1}, \cref{table:errorinf} and \cref{table:errordistribution}).
	
	\begin{figure}[!ht]
		\centering
		\begin{subfigure}[b]{0.75\textwidth}
			\includegraphics[width=1\textwidth]{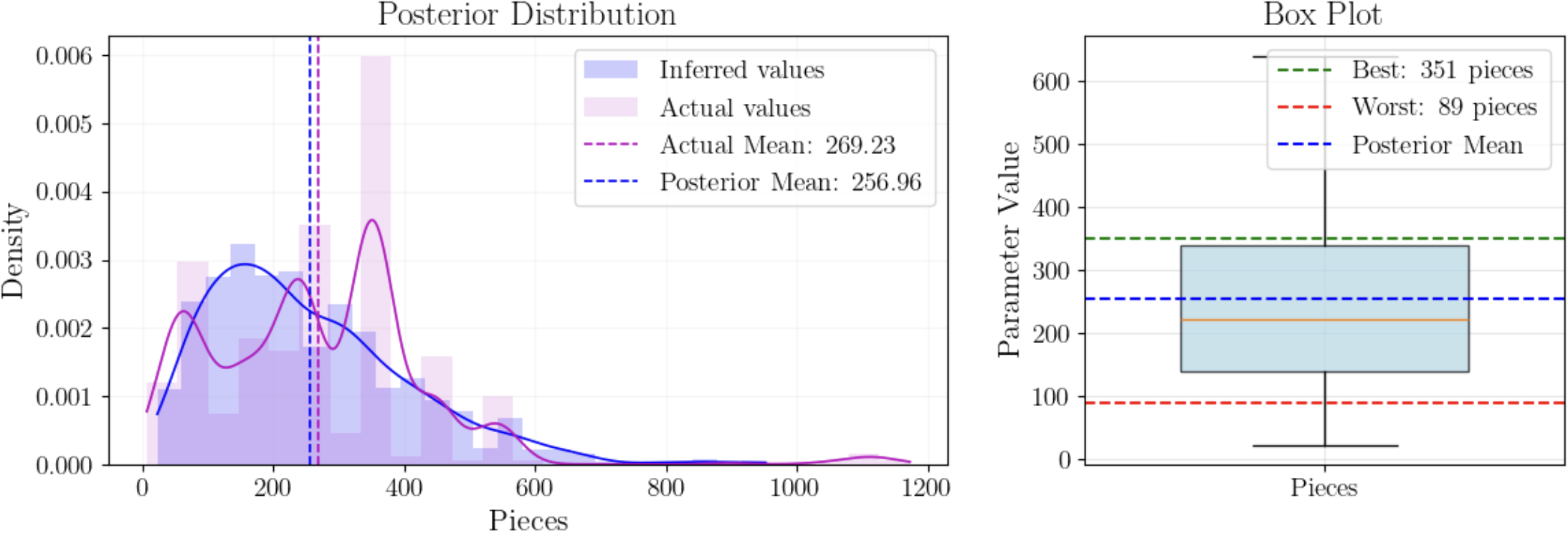}
			\caption{}\label{fig:InvBa}
		\end{subfigure}
		\begin{subfigure}[b]{0.75\textwidth}
			\includegraphics[width=1\textwidth]{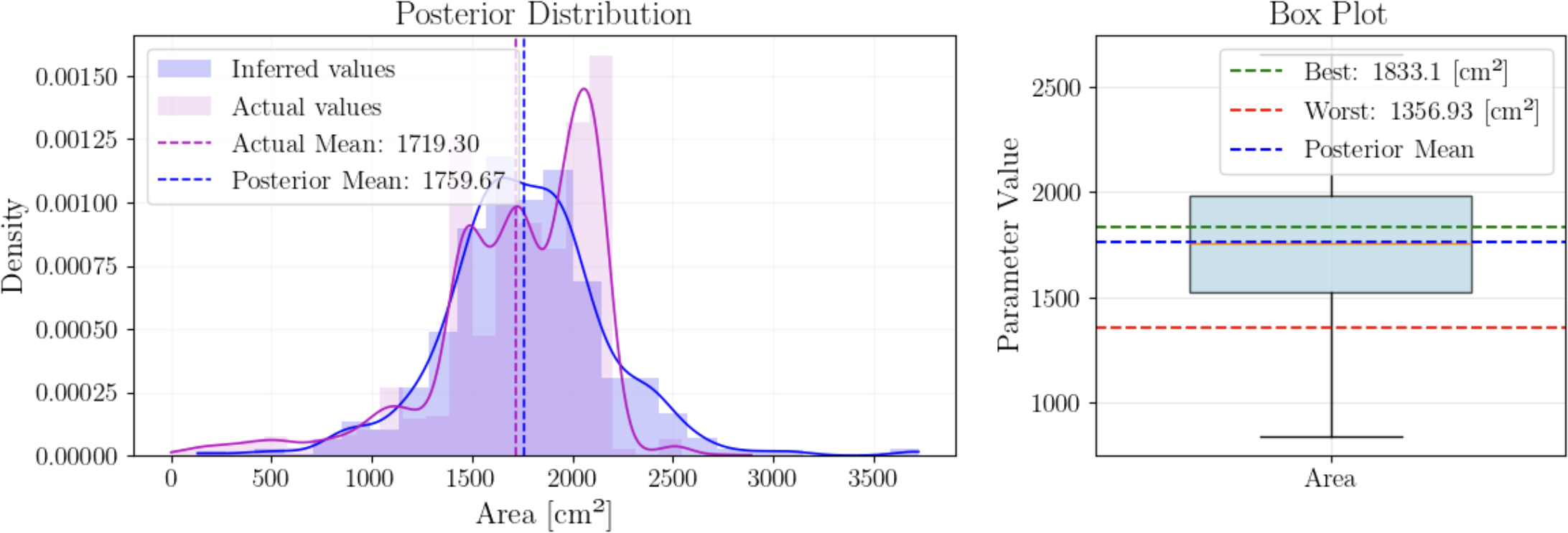}
			\caption{}\label{fig:InvBb}
		\end{subfigure}
		\begin{subfigure}[b]{0.75\textwidth}
			\includegraphics[width=1\textwidth]{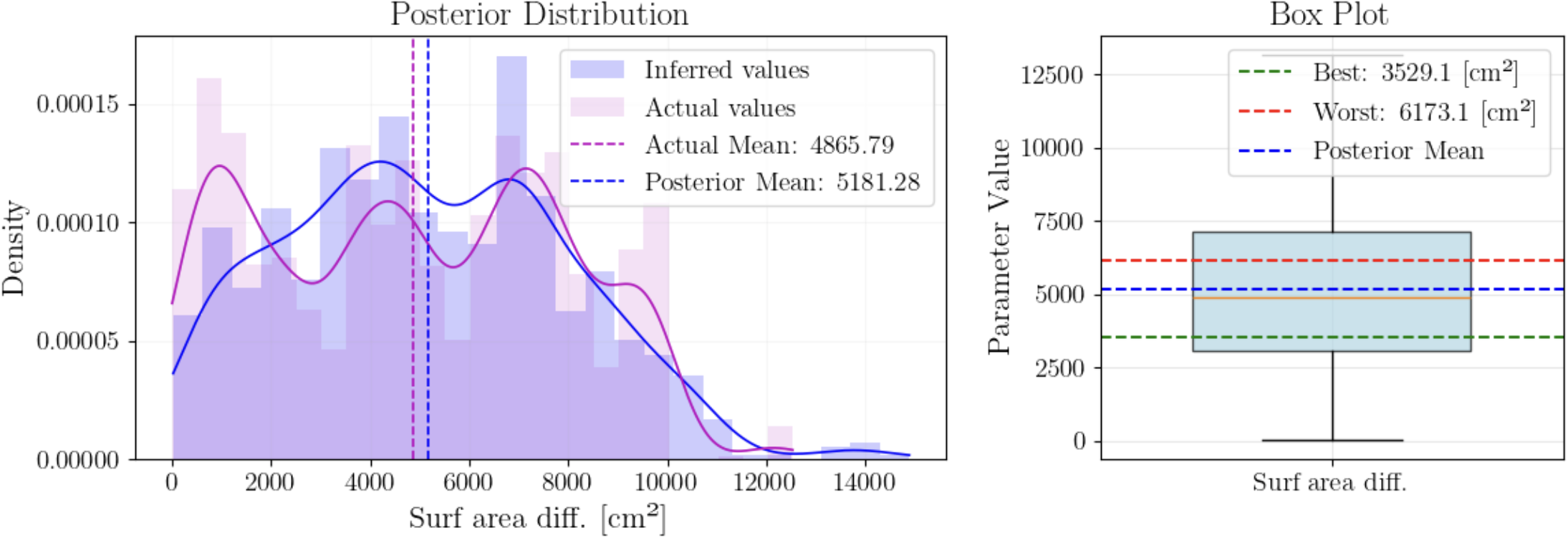}
			\caption{}\label{fig:InvBc}
		\end{subfigure}
		\caption{Posterior distributions and credible intervals for key process parameters inferred using weighted ABC sampling (a) for "Pieces", (b) for "Area", and (c) for "Surface area difference". Left figures show kernel density estimates comparing inferred posterior distributions (blue) with actual parameter distributions from historical data (violet). Dashed vertical lines indicate their corresponding means. Right figures display boxplot-style credible intervals showing the 95\% CI (outer boundaries), interquartile range/50\% CI (inner box, light blue), posterior median (orange line), and posterior mean (blue line). Green and red dashed horizontal lines indicate best and worst production run parameter values, respectively.}
		\label{fig:InvB}
	\end{figure}
	
	The logistic error distribution defines the kernel function $$\mathcal{K}(d; \text{Logistic}(\mu, s)) = \dfrac{1}{s \left(1 + \exp\left(-\dfrac{d - \mu}{s}\right)\right)^2},$$ which computes importance weights to identify the most representative samples of the posterior $\pi(\bm{x}|y_\text{obs})$.
	
	The ABC algorithm successfully generates posterior samples for reactor setup parameters $\bm{x}$ , given observed mean coating thickness measurements $y_\text{obs}$. These initial results correspond to the Bayesian implementation using a predictor trained on data incorporating binary embeddings for categorical parameters, achieving an effective sample size (ESS) of 99.73\% from 1000 generated samples, indicating adequate posterior representation. 
	
	In particular, due to their interpretability and relevance to the coating process, as suggested by expert knowledge and confirmed by the feature importance analysis in \cref{fig:IMPp}, we focus on analyzing three of the most significant setup parameters: "Pieces", "Area", and "Surface Area Difference".
	
	For the parameter "Pieces" (\cref{fig:InvBa}),  the posterior mean is 257 pieces closely approximates the actual mean of 269, with a 95\% credibility interval spanning from approximately 45 to 670 pieces. The posterior distribution assign substantial probability mass to values near the best observed production run performance (351 pieces), suggesting the inference successfully identified parameter regions associated with optimal outcomes. 
	In practical terms, these inferred values could help establish an appropriate number of pieces to be loaded onto each tray depending on the size or shape of the inserts (see \cref{Subsec:Geometry}).
	
	The "Area" parameter demonstrates better agreement between inferred and actual distributions (\cref{fig:InvBb}). The posterior mean is 1759.67 cm$^2$, indicating high accuracy in parameter probabilistic description. The 95\% credibility interval (1038.04, 2494.66 cm$^2$) covers the best observed value of 1833.1 cm$^2$, though the posterior assigns higher probability density to slightly lower values. According to expert insight, a larger coated area leads to produce a more uniform coating thickness. This trend is verified by these findings, as the inferred values are concentrated above those corresponding to the least efficient production run.
	
	For "Surface area difference", the posterior distribution (\cref{fig:InvBc}). The posterior mean is 5181.28 cm$^2$ that slightly exceeds the actual mean of 4865.79 cm$^2$, with the 95\% credibility interval ranging from 425.14 to 10908.21 cm$^2$. Both the best (3529.1 cm$^2$) and worst (6173.1 cm$^2$) observed values fall within the posterior high-probability region, indicating the inference captured the parameter range relevant to process performance variation. Furthermore, expert knowledge indicates that this parameter, defined as the difference between the nominal surface area (from the recipe) and the actual surface area loaded into the reactor, should be as small as possible in order to achieve the desired coating uniformity. The obtained results align with this principle, since the mean of the inferred values is lower than the one observed in the worst production run.
	
	\begin{figure}[!ht]
		\centering
		\includegraphics[width=.55\linewidth]{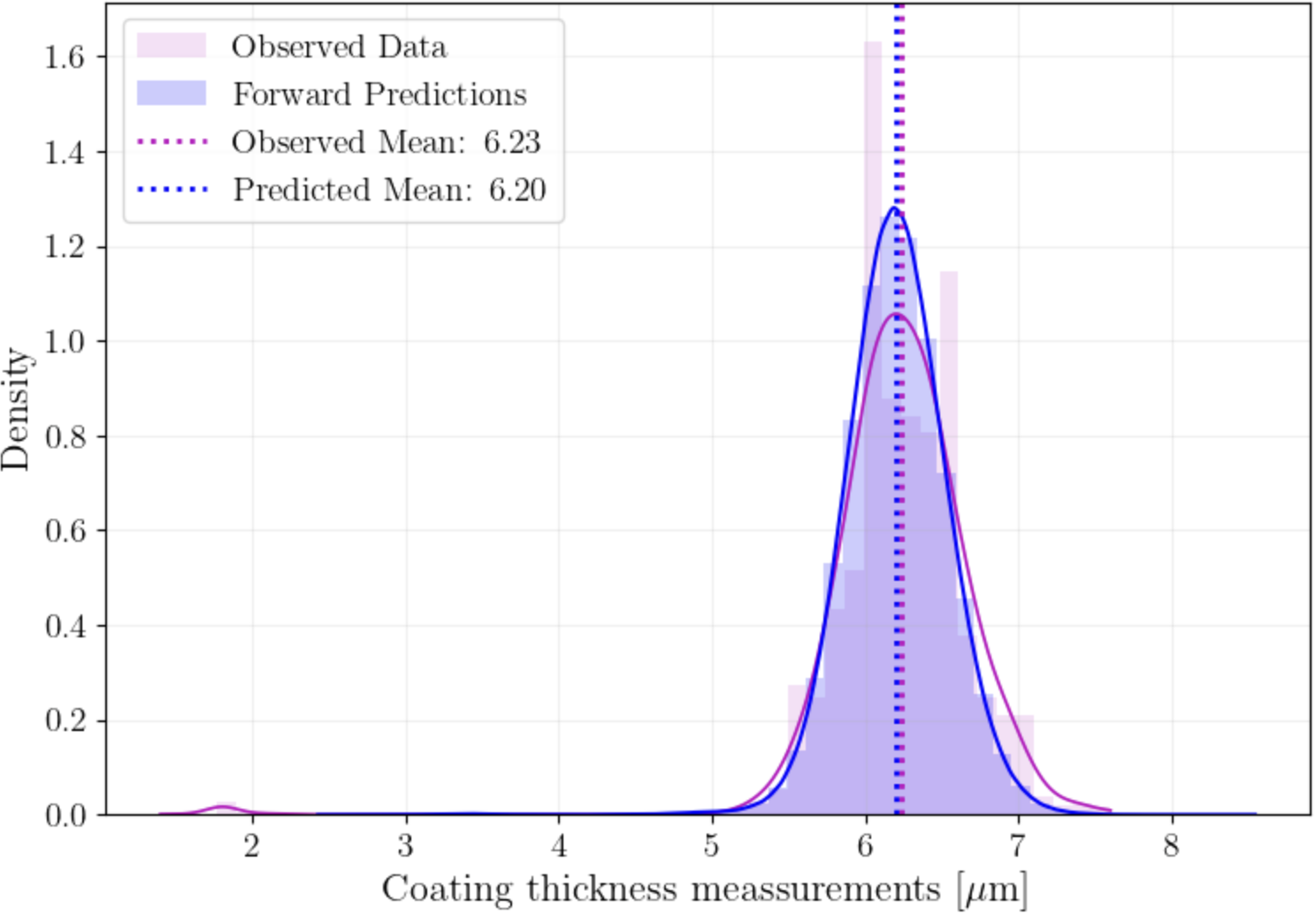}
		\caption{Forward validation comparing observed process data with predictions generated from posterior parameter samples. Histograms and kernel density estimates show the distribution of insert thickness mean for observed data (orange) and forward predictions using ABC-inferred parameters (blue). Vertical dashed lines indicate distribution means.}
		\label{fig:outB}
	\end{figure}
	
	Forward validation (\cref{fig:outB}) demonstrates excellent agreement between observed data and predictions generated from posterior parameter samples. The kernel density estimates of both distributions show a mean difference of 0.03 $\mu$m (observed: 6.23 $\mu$m vs. predicted: 6.20 $\mu$m). 
	
	These results confirm that the weighted ABC approach effectively captured the underlying parameter relationships, despite the discrepancies observed in the individual parameter posteriors. In addition, the inferred parameters seem to mitigate some of the extreme cases, converging toward more plausible values around the mean.

	\subsection{Shape-based Clustering and ABC Inference}\label{Subsec:Geometry}
	
\begin{figure}[!ht]
		\centering
		\includegraphics[width=0.45\linewidth]{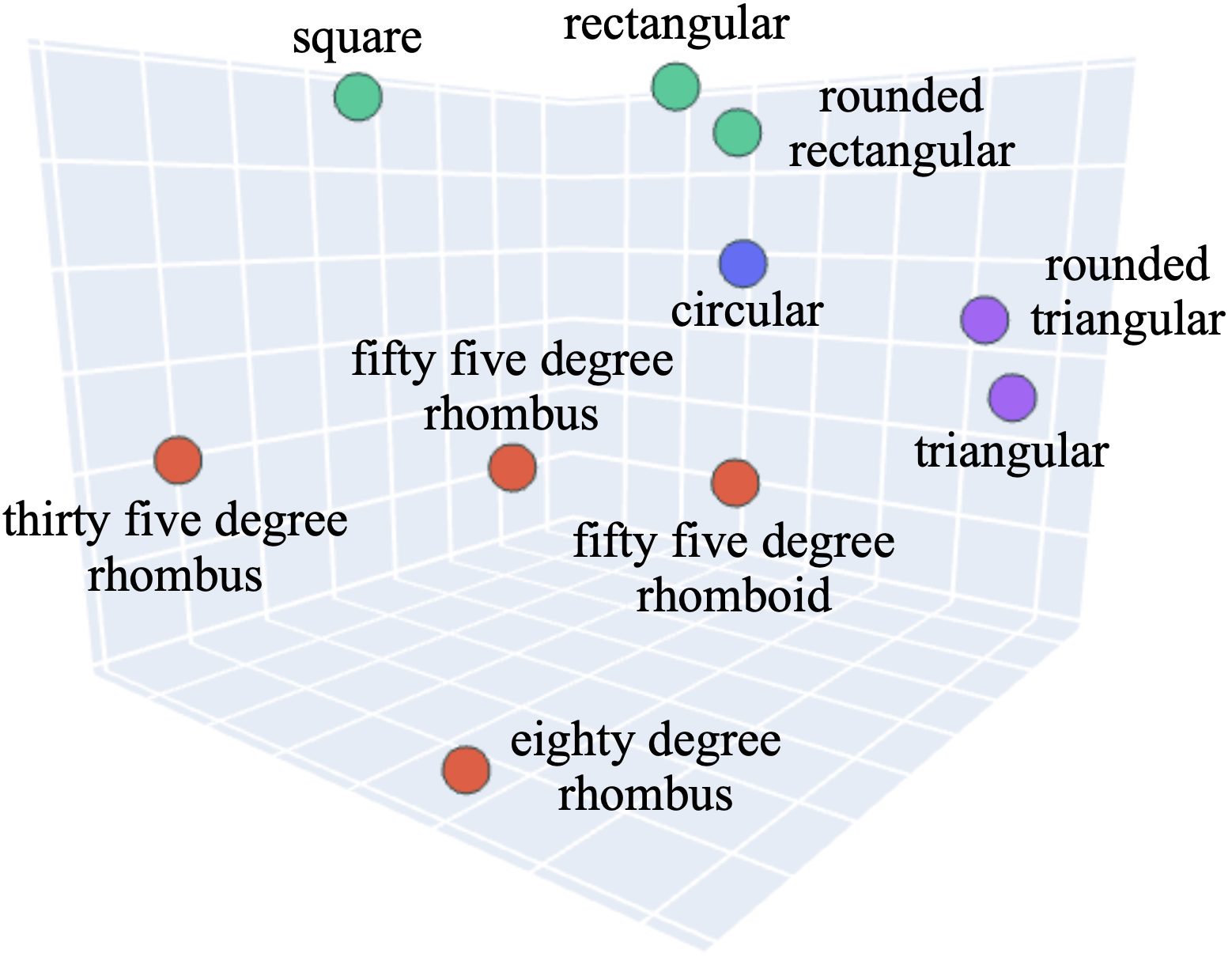}
		\caption{The 3D representation of the embeddings associated with the insert shapes reveals four geometrically homogeneous clusters corresponding to triangular (purple), rhomboid (red), circular (blue), and rectangular (green) insert families.}
		\label{fig:clus}
	\end{figure}

	To enable a more refined analysis of the relationship between insert geometries and process performance, we implemented a shape representation strategy using text embeddings. Unlike traditional binary encoding approaches that treat shape categories as independent variables, our method leveraged natural language descriptions of insert geometries to generate dense embedding vectors (dimension = 3, cf. \cref{fig:clus}). Pairwise cosine similarities between these embeddings revealed meaningful geometric relationships (see \cref{fig:cosine}). Spectral clustering based on these similarity metrics produced four distinct, geometrically homogeneous clusters corresponding to triangular (inner similarity > 0.98), rhomboid (inner similarity > 0.61), circular, and rectangular (inner similarity > 0.85) insert families, which were subsequently used to stratify the ABC inference. 
	
	\begin{figure}[!ht]
		\centering
		\includegraphics[width=0.7\linewidth]{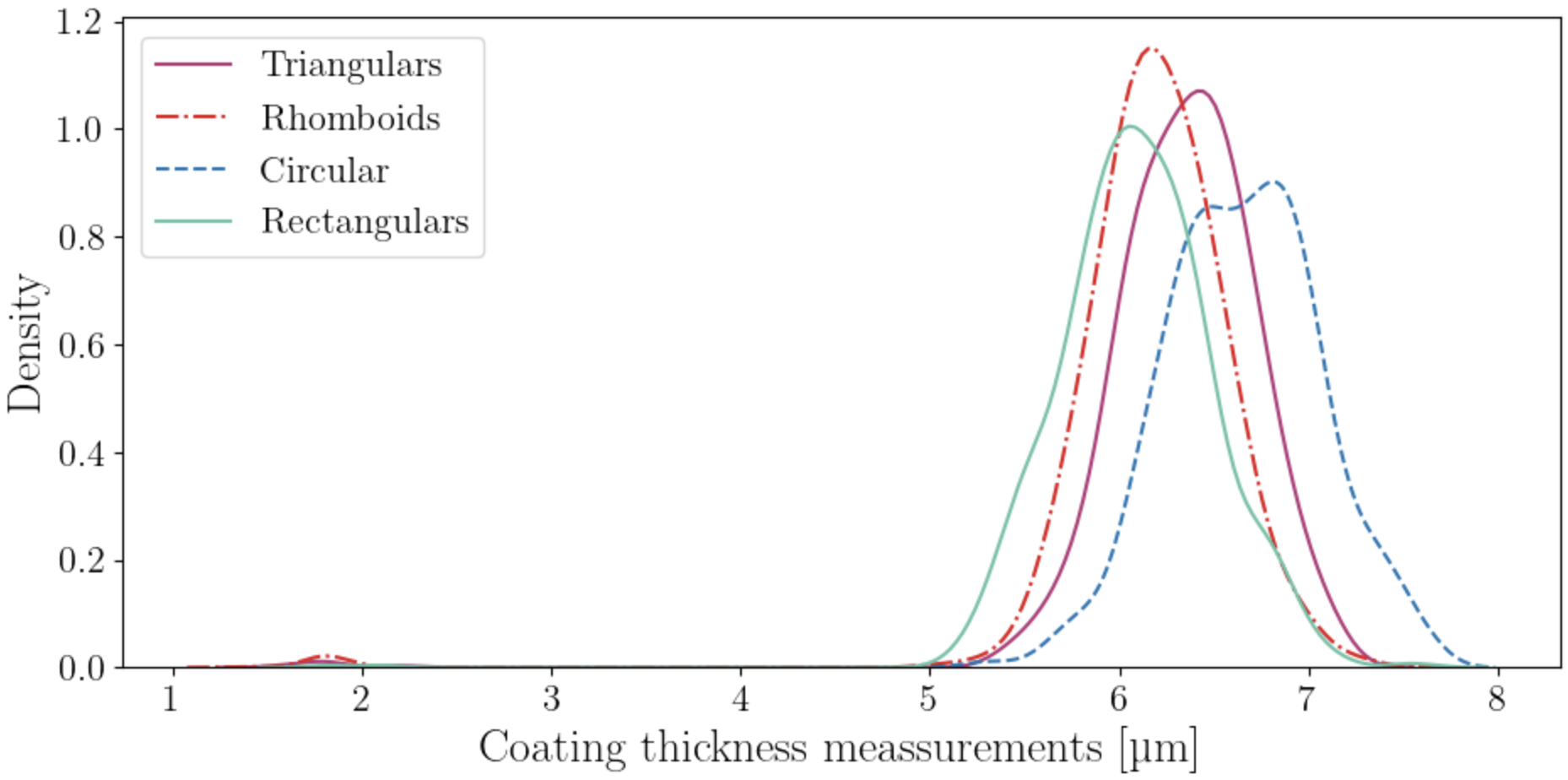}
		\caption{Coating thickness distributions for each geometric cluster (triangular, rhomboid, circular and rectangular) showing distinct process response behaviors.}
		\label{fig:clusimg}
	\end{figure}
	
	\begin{table}[!ht]
		\centering
		\caption{Cluster composition following shape-based stratification. Each cluster represents geometrically similar insert variants, with sample size and their relative proportion of the total dataset (dimension=3635).}
		\begin{tabular}{ccc}
			\hline
			\textbf{Shape} & \textbf{Size} & \textbf{Percentage [\%]} \\
			\hline
			Triangular & 973 & 26.77 \\
			Rhomboid & 2085 & 57.36 \\
			Circular & 293 & 8.06 \\
			Rectangular & 827 & 22.75 \\
			\hline
		\end{tabular}
		\label{table:clusters}
	\end{table}
	
	\textbf{Triangular Inserts:} Triangular insert variants constituted 26.77\% of the dataset (see \cref{fig:clusimg} and \cref{table:clusters}). Posterior inference for this cluster yielded narrow credible intervals indicating low process variability: "Area" posterior mean of 1520.70 cm$^2$ (95\% CI: 1152.14-2047.82 cm$^2$), "Pieces" mean of 268 units (95\% CI: 98-500), and Surface area difference achieving 5992.85 cm$^2$ (95\% CI: 346.72-11910.14 cm$^2$).
	
	The coating thickness posterior (mean = 6.32 $\mu$m, SD = 0.32 $\mu$m) showed a right-shift relative to empirical observations with reduced variance, suggesting that triangular geometries enable tighter process control and consistently achieve the coating uniformity required for high production quality (see Appendix \ref{Appendix2}).
	
	\textbf{Rhomboid Insert} Cluster encompassed all diamond-shaped inserts including 80°, 55°, and 35° rhombic variants, representing 57.36\% of the dataset (see \cref{table:clusters}). This cluster exhibited intermediate posterior uncertainty, with "Area" spanning 1240.81-3444.45 cm$^2$ (posterior mean: 2117.70 cm$^2$), number of "Pieces" between 94 and 731 (posterior mean: 329 pieces), and "Surf area diff." showing 95\% CI of 425.97-12228.29 cm$^2$. The posterior distribution for "Insert thickness" displayed a bimodal structure, suggesting that different rhomboid variants operate optimally under distinct thickness regimes. The coating thickness posterior (mean = 6.10 $\mu$m, SD = 0.35 $\mu$m) showed lower values, suggesting uniformity required for high production quality (see Appendix \ref{Appendix3}).

	\textbf{Circular Insert:} Cluster comprised all round insert variants represent 8.06\% of the dataset (see \cref{table:clusters}). The posterior distribution for this cluster showed the tightest credibility intervals among all geometries, with the "Area" parameter converging to 1660.55 cm$^2$ (95\% CI: 1097.84-2280.41 cm$^2$), "Pieces" achieving 86 units (95\% CI: 2-337 units), and "Surf area diff." showing 95\% CI of 309.39-9729.05 cm$^2$. The narrow posteriors reflect reduced process variability associated with circular geometries, likely due to their continuous cutting edges and absence of corner wear concentration. Forward validation for this cluster demonstrates that the coating thickness posterior (mean = 6.57 $\mu$m, SD = 0.35 $\mu$m) is right-shifted relative to observed data with reduced variance, indicating that triangular inserts achieve both greater mean coating thickness and lower variability (see Appendix \ref{Appendix4}).
	
	\textbf{Rectangular Insert} Cluster grouped all square and rectangular insert variants with various edge preparations, representing 232.75\% of the dataset (see \cref{table:clusters}) The ABC inference for this cluster reveals high parameter uncertainty, with "Area" 95\% CI spanning 680.87-2393.46 cm$^2$ (posterior mean: 1481.01 cm$^2$) and "Surf area diff." ranging from 755.54 to 10796.02 cm$^2$. This substantial uncertainty reflects the inherent geometric discontinuities at corners in rectangular inserts, which create less predictable cutting conditions and higher process variability. The posterior for "Pieces" (143.01 units, 95\% CI: 3-516) is lower and more dispersed than the other geometries. Despite shape-related uncertainties, the process consistently delivers high production quality, reflected in the narrow coating thickness posterior (mean = 6.21 $\mu$m, SD = 0.36 $\mu$m, see Appendix \ref{Appendix5}).
	
	Comparison of effective sample sizes across shape families reveals that circular inserts achieved the highest effective sample size (ESS$_{\text{circular}} = 99.78\%$), followed by rhomboid (ESS$_{\text{rhomboid}} = 99.64\%$), triangular (ESS$_{\text{triangular}} = 99.63\%$), and rectangular (ESS$_\text{rectangular} = 99.61\%$), confirming progressively more concentrated posterior mass and reduced parameter uncertainty for geometrically smoother shapes.
	
	This shape-based stratification was only possible because the embedding representation preserved geometric family structure through semantic similarity. In binary encoding, categorical parameters are encoded as completely independent categories with no encoded relationship, preventing any clustering algorithm from recognizing they belong to the same geometric family. The embedding approach encoded the shared geometric descriptor "circular" across all variants. This fundamental difference, preservation of geometric family structure, makes embedding-based stratified inference uniquely capable of revealing shape-specific parameter-performance relationships that remain hidden binary-encoded analyses.

	\section{Conclusions}\label{sec:Conclusions}
	
	This work demonstrates the integration of Approximate Bayesian Computation with XGBoost surrogate modeling and binary/\textit{Doc2Vec} embeddings for inverse uncertainty quantification in industrial CVD processes. The proposed methodology achieves computational efficiency through likelihood-free inference while aligning with expert insights, offering significant speed advantages over traditional MCMC approaches without sacrificing accuracy. The framework exhibits scalability when handling mixed-type parameter spaces, effectively managing both continuous and categorical parameters such as reactor configurations and insert geometries. The use of feature importance analysis provides clear interpretability of parameter influence, enabling engineers to understand which factors most significantly impact coating uniformity while quantifying associated uncertainties.
	
	The integration of \textit{Doc2Vec} embeddings for categorical variable representation in industrial process modeling is leveraged in this application and demonstrates how NLP techniques can effectively encode geometric specifications and enable the implementation of process optimization using similarity assessment and parameter grouping compared to traditional categorical encoding methods that are not able to captures similarities between categories. This allow us to focus the analysis on particular parameters such us insert geometries.
	
	The practical validation using actual production data demonstrates the applicability of this framework to industrial settings. Additionally, it allows us to infer meaningful parameter distributions from noisy industrial data addresses a critical need in manufacturing environments where comprehensive experimental design is often prohibitively expensive. By leveraging existing production data, the methodology reduces the experimental cost typically required for robust parameter estimation while providing uncertainty quantification essential for reliable industrial decision-making.
	
	The demonstrated success in CVD process optimization indicates broader applicability across chemical manufacturing, materials processing, and quality control applications. This methodology is able to handle mixed-type parameters and provide uncertainty quantification makes it particularly valuable for complex industrial systems where traditional optimization methods fall short. The methodology emphasis on using actual production data aligns with Industry initiatives focusing on data-driven process improvement and predictive analytics.
	
	This work contributes to the growing intersection of Bayesian inference and machine learning in industrial applications, providing a new and practical template for implementing advanced uncertainty quantification methods in real-world manufacturing environments. The validation framework developed here offers a systematic approach for assessing estimation quality in industrial settings, addressing a critical gap in the practical application of Bayesian methods to manufacturing processes.
	
	Future research should focus on adaptive ABC implementations with dynamic tolerance and summary statistics selection to reduce the need for domain expertise and improve automation. Extension to multi-output scenarios would enhance the framework applicability to processes with multiple correlated quality metrics. Real-time implementation strategies would enable online estimation with streaming data.

	\section*{Acknowledgments}
	
	This research was funded in part by the Luxembourg National Research Fund (FNR), grant reference [16758846]. For the purpose of open access, the authors have applied a Creative Commons Attribution 4.0 International (CC BY 4.0) license to any Author Accepted Manuscript version arising from this submission. GLS gratefully acknowledges funding from the FSTM in the University of Luxembourg.

	\bibliographystyle{elsarticle-num-names} 
	\bibliography{references}

	\newpage
	\begin{appendices}
		\crefalias{section}{appsec}
		\section{Appendix}\label{Appendix}
		
		\subsection{Residual distributions}\label{Appendix1}
		
		\begin{figure}[!ht]
			\centering
			\begin{subfigure}[b]{0.85\textwidth}
				\includegraphics[width=1\textwidth]{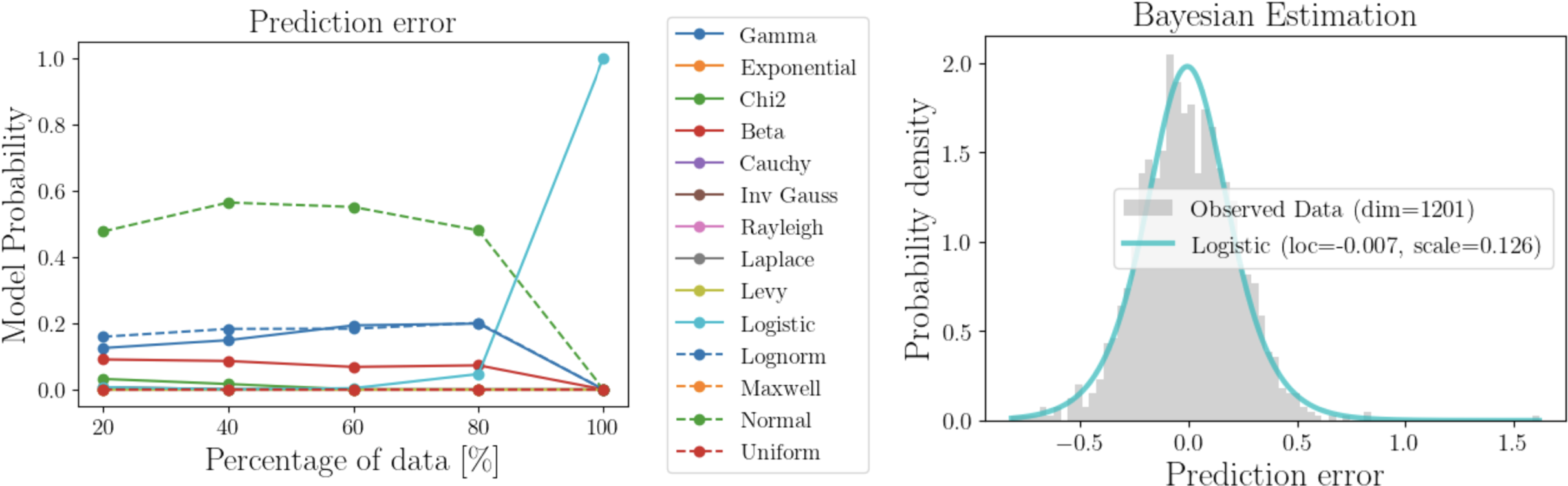}
				\caption{}
			\end{subfigure}
			\begin{subfigure}[b]{0.85\textwidth}
				\includegraphics[width=1\textwidth]{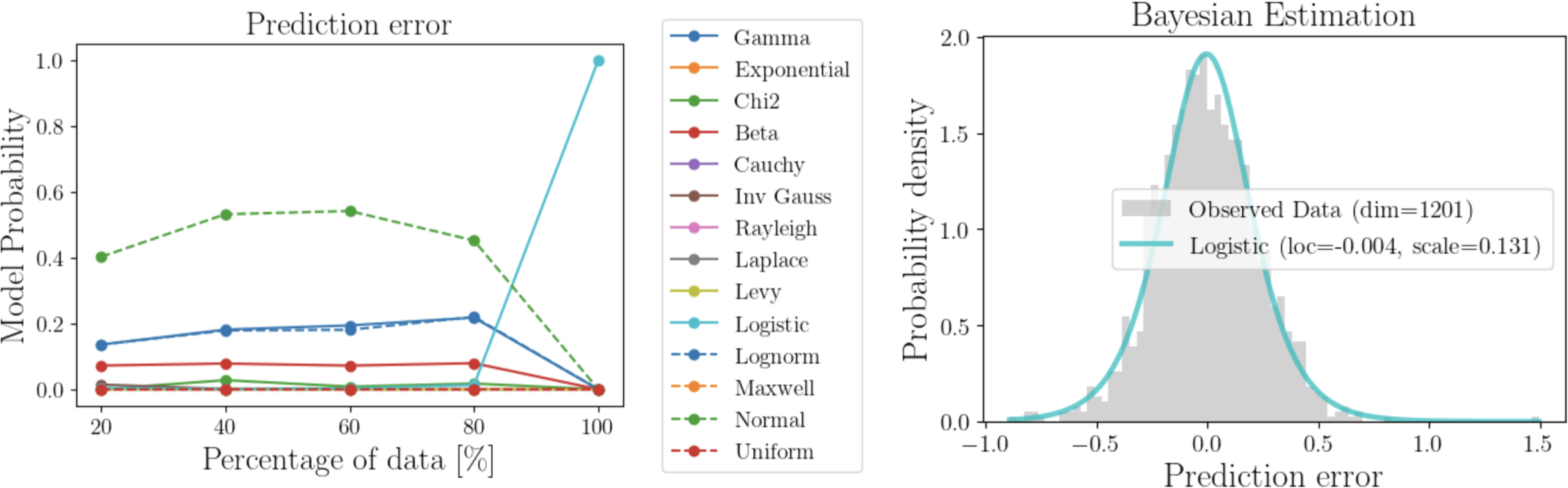}
				\caption{}
			\end{subfigure}
			\caption{Model selection and parameter estimation results for prediction error (residuals). Left: Model probabilities computed using AIC. Right: Posterior distributions of the best-fitting models. (a) For the predictor model with binary encoding and (b) for the predictor model with \textit{Doc2Vec} embeddings, shown alongside observed data histograms (gray bars), demonstrating good model-data agreement.}\label{table:errorinf}
		\end{figure}
		\begin{table}[ht]
			\centering
			\caption{Inferred distributions for prediction error (residuals) estimated via Bayesian Parameter Estimation. Distribution types were selected using AIC, and hyperparameters represent MCMC posterior means.}
			\begin{tabular}{ccc}
				\hline
				\textbf{Prediction error}                                                                           & \textbf{Probability Distribution}       & \textbf{Hyperparameters} \\
				\hline
				Predictor with binary encoding & Logistic & $\mu = -0.007,\ s = 0.126$\vspace{0.15 cm}        \\
				Predictor with \textit{Doc2Vec} embedding & Logistic& $\mu = -0.004,\ s = 0.131$ \\
				\hline
			\end{tabular}
			\label{table:errordistribution}
		\end{table}

		\newpage
		\subsection{Posterior distributions for triangular insert shapes }\label{Appendix2}
		\begin{figure}[!ht]
			\centering
			\begin{subfigure}[b]{0.65\textwidth}
				\includegraphics[width=1\textwidth]{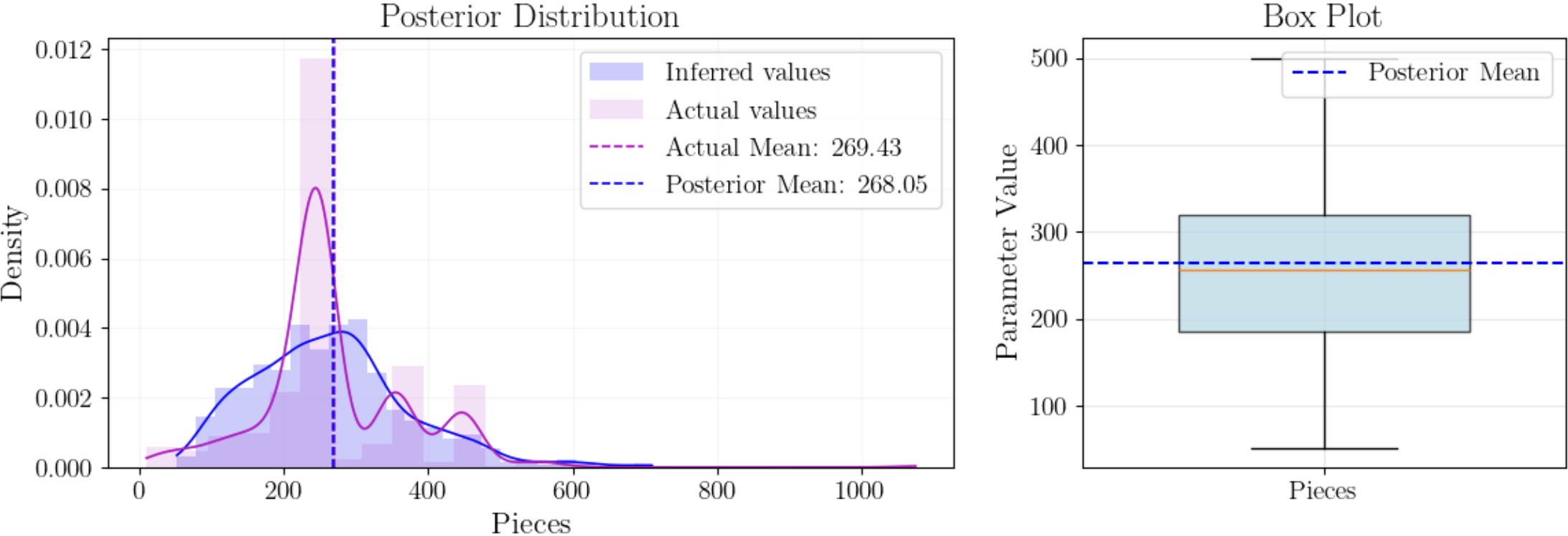}
				\caption{}
			\end{subfigure}
			\begin{subfigure}[b]{0.65\textwidth}
				\includegraphics[width=1\textwidth]{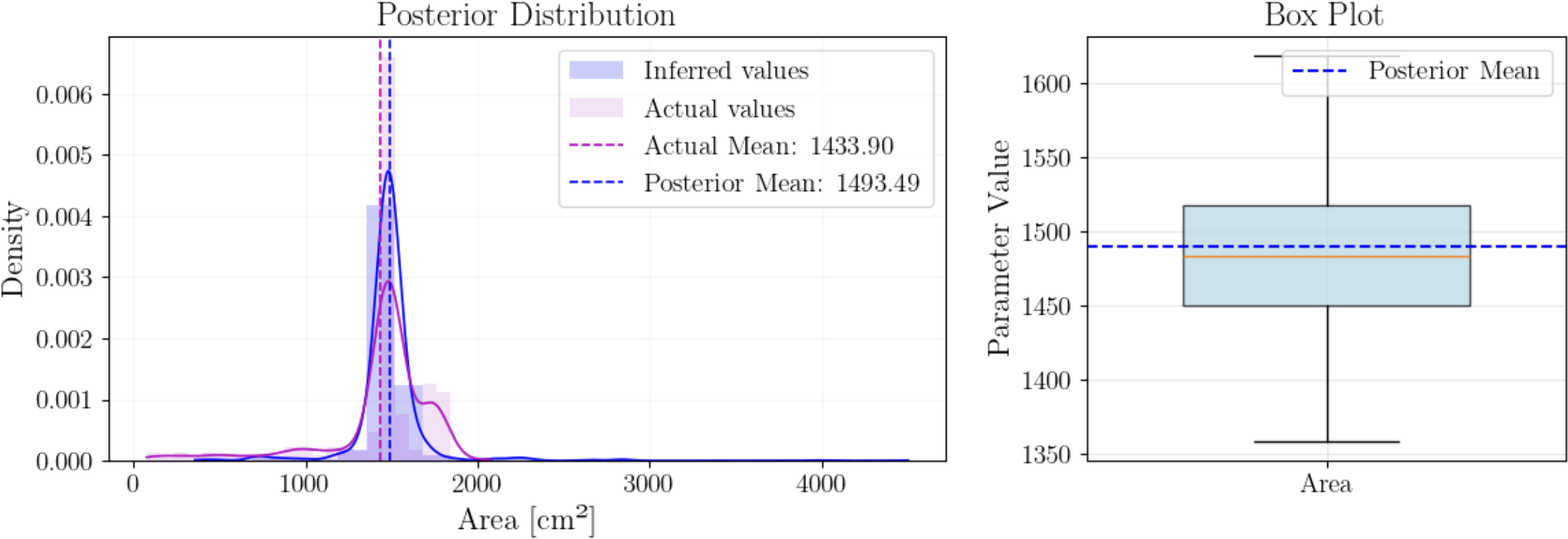}
				\caption{}
			\end{subfigure}
			\begin{subfigure}[b]{0.65\textwidth}
				\includegraphics[width=1\textwidth]{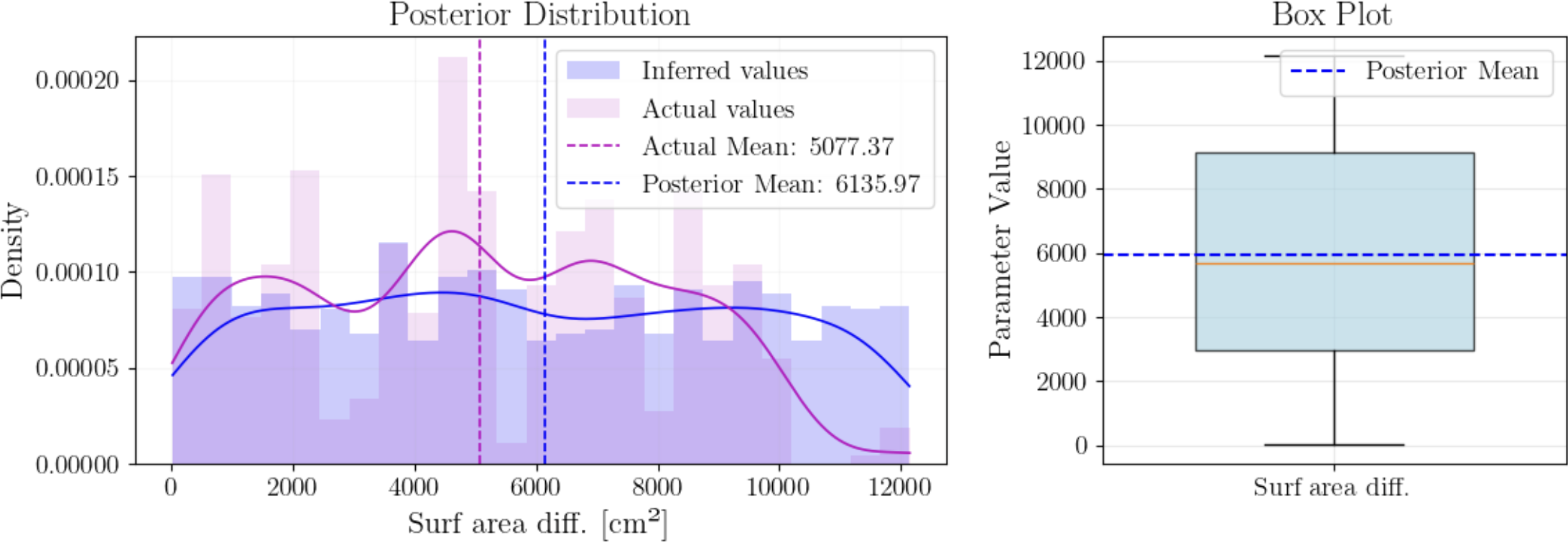}
				\caption{}
			\end{subfigure}
			\caption{For triangular insert shapes: posterior distributions and credible intervals for key process parameters inferred using weighted ABC sampling (a) for "Pieces", (b) for "Area", and (c) for "Surface area difference". Left figures show kernel density estimates comparing inferred posterior distributions (blue) with actual parameter distributions from historical data (violet). Dashed vertical lines indicate their corresponding means. Right figures display boxplot-style credible intervals showing the 95\% CI (outer boundaries), 50\% CI (inner box, light blue), posterior median (orange line), and posterior mean (blue line).}
			\label{fig:Invtri}
		\end{figure}
		\begin{figure}[!ht]
			\centering
			\includegraphics[width=.4\linewidth]{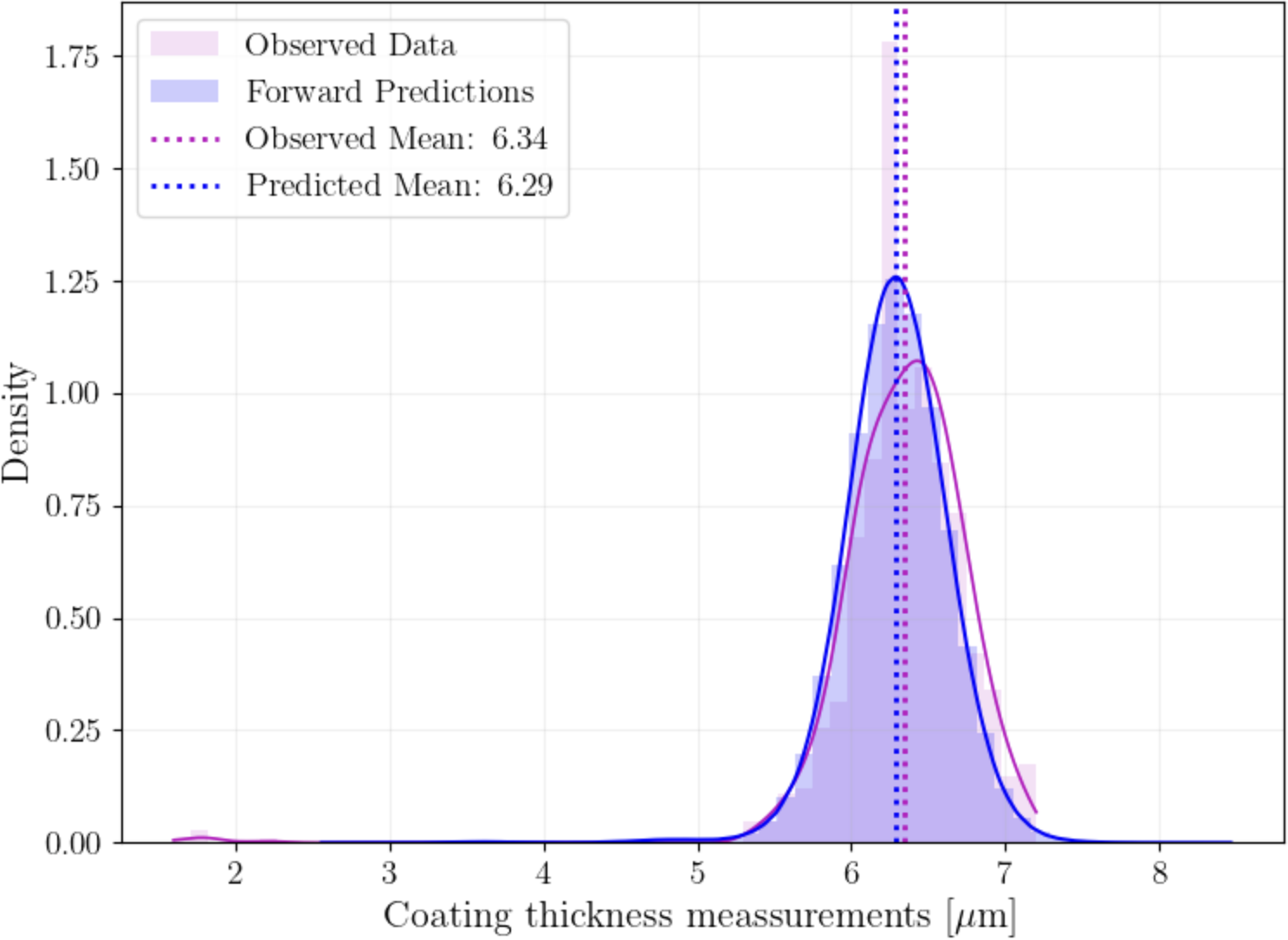}
			\caption{For triangular insert shapes:: forward validation comparing observed process data with predictions generated from posterior parameter samples. Histograms and kernel density estimates show the distribution of insert thickness for observed data (orange) and forward predictions using ABC-inferred parameters (blue). Vertical dashed lines indicate distribution means.}
			\label{fig:outtri}
		\end{figure}

		\newpage
		\subsection{Posterior distributions for rhomboid insert shapes }\label{Appendix3}
		\begin{figure}[!ht]
			\centering
			\begin{subfigure}[b]{0.65\textwidth}
				\includegraphics[width=1\textwidth]{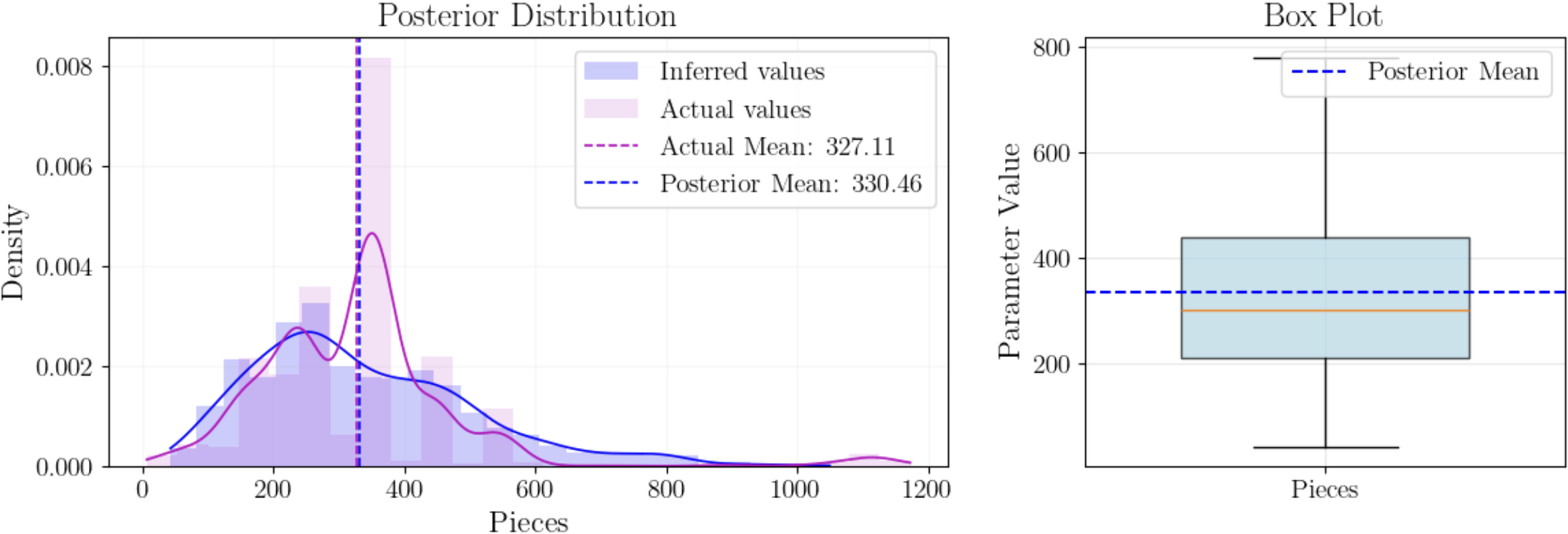}
				\caption{}
			\end{subfigure}
			\begin{subfigure}[b]{0.65\textwidth}
				\includegraphics[width=1\textwidth]{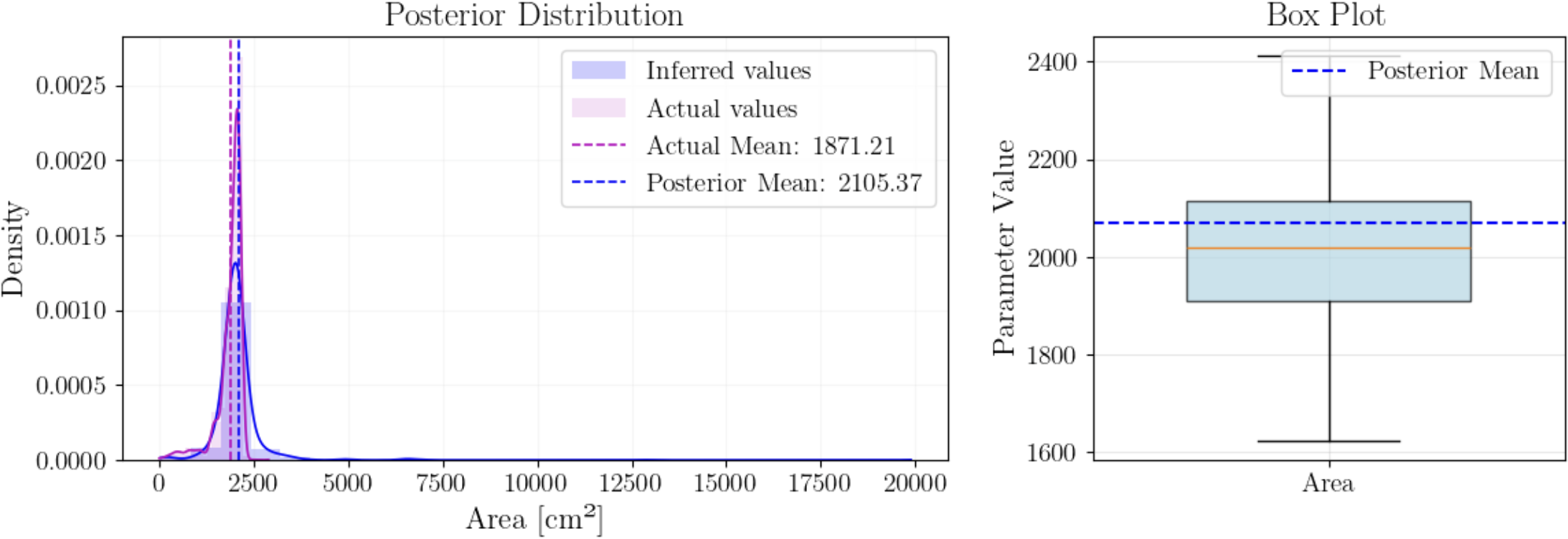}
				\caption{}
			\end{subfigure}
			\begin{subfigure}[b]{0.65\textwidth}
				\includegraphics[width=1\textwidth]{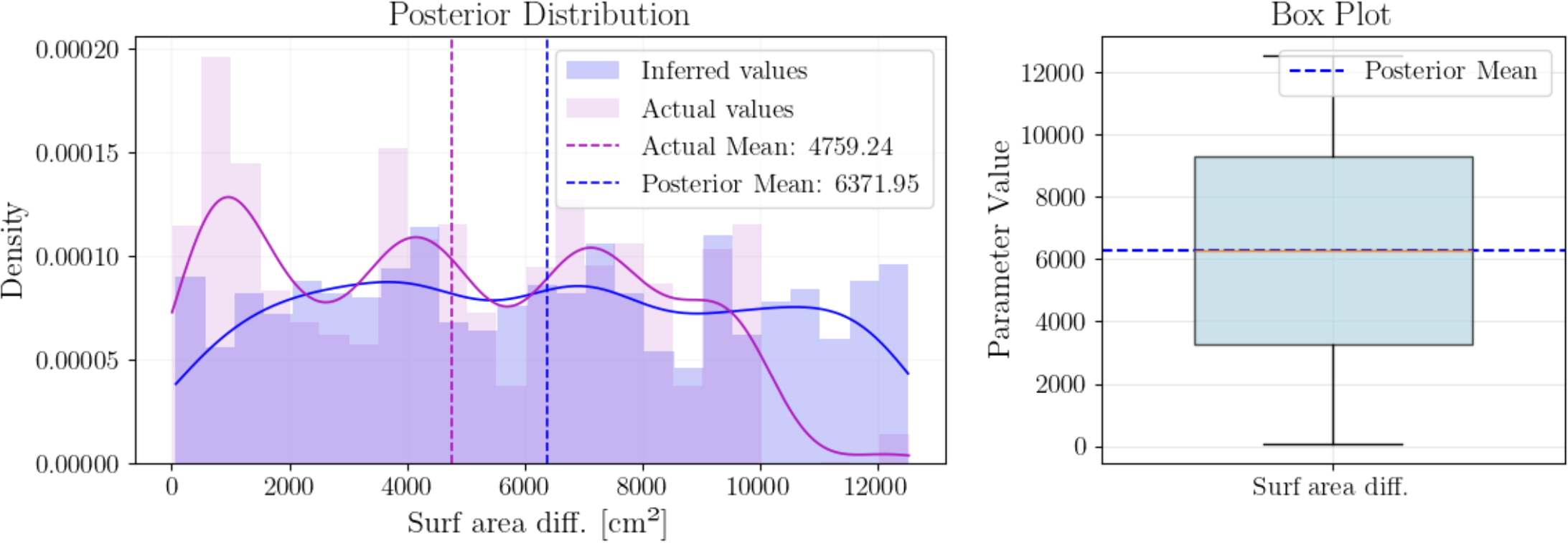}
				\caption{}
			\end{subfigure}
			\caption{For rhomboid insert shapes: posterior distributions and credible intervals for key process parameters inferred using weighted ABC sampling (a) for "Pieces", (b) for "Area", and (c) for "Surface area difference". Left figures show kernel density estimates comparing inferred posterior distributions (blue) with actual parameter distributions from historical data (violet). Dashed vertical lines indicate their corresponding means. Right figures display boxplot-style credible intervals showing the 95\% CI (outer boundaries), 50\% CI (inner box, light blue), posterior median (orange line), and posterior mean (blue line).}
			\label{fig:Invrhom}
		\end{figure}
		
		\begin{figure}[!ht]
			\centering
			\includegraphics[width=.4\linewidth]{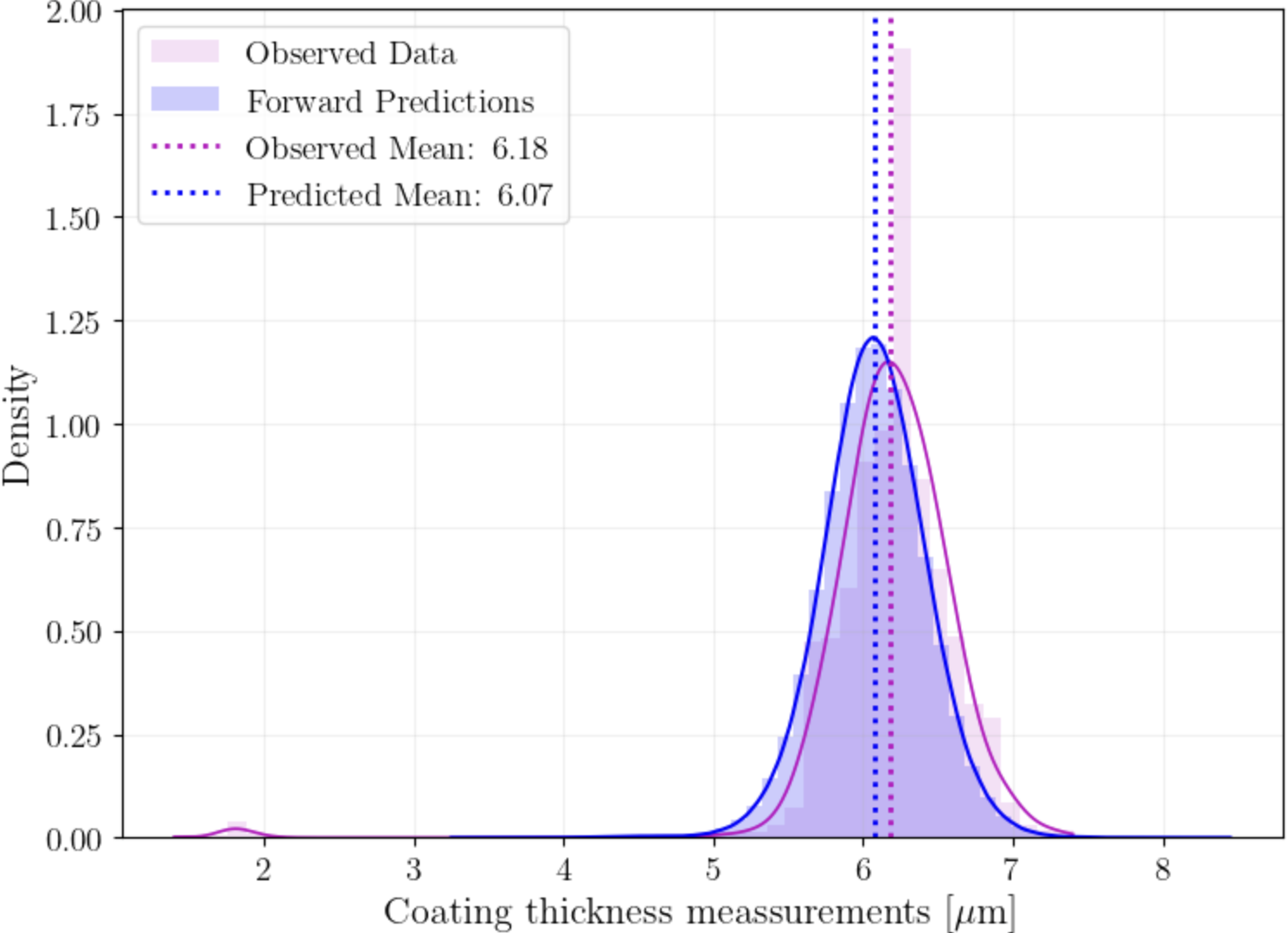}
			\caption{For rhomboid insert shapes:: forward validation comparing observed process data with predictions generated from posterior parameter samples. Histograms and kernel density estimates show the distribution of insert thickness for observed data (orange) and forward predictions using ABC-inferred parameters (blue). Vertical dashed lines indicate distribution means.}
			\label{fig:outrhom}
		\end{figure}

		\newpage
		\subsection{Posterior distributions for circular insert shapes }\label{Appendix4}
		\begin{figure}[!ht]
			\centering
			\begin{subfigure}[b]{0.65\textwidth}
				\includegraphics[width=1\textwidth]{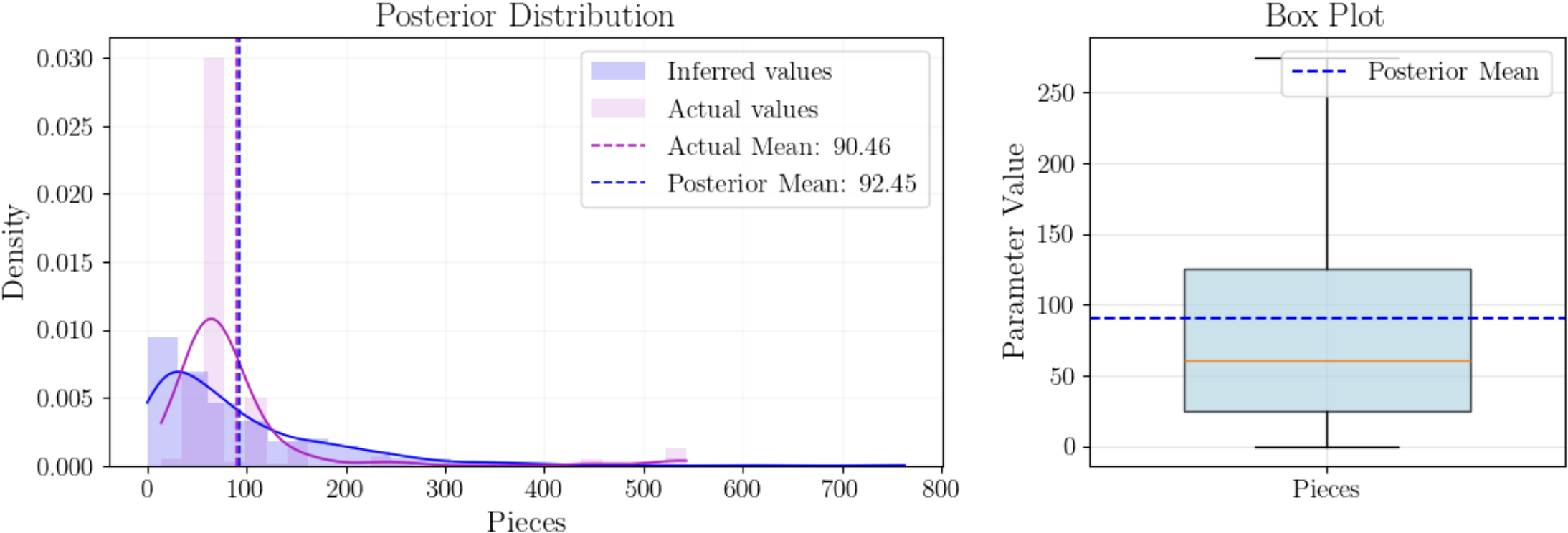}
				\caption{}
			\end{subfigure}
			\begin{subfigure}[b]{0.65\textwidth}
				\includegraphics[width=1\textwidth]{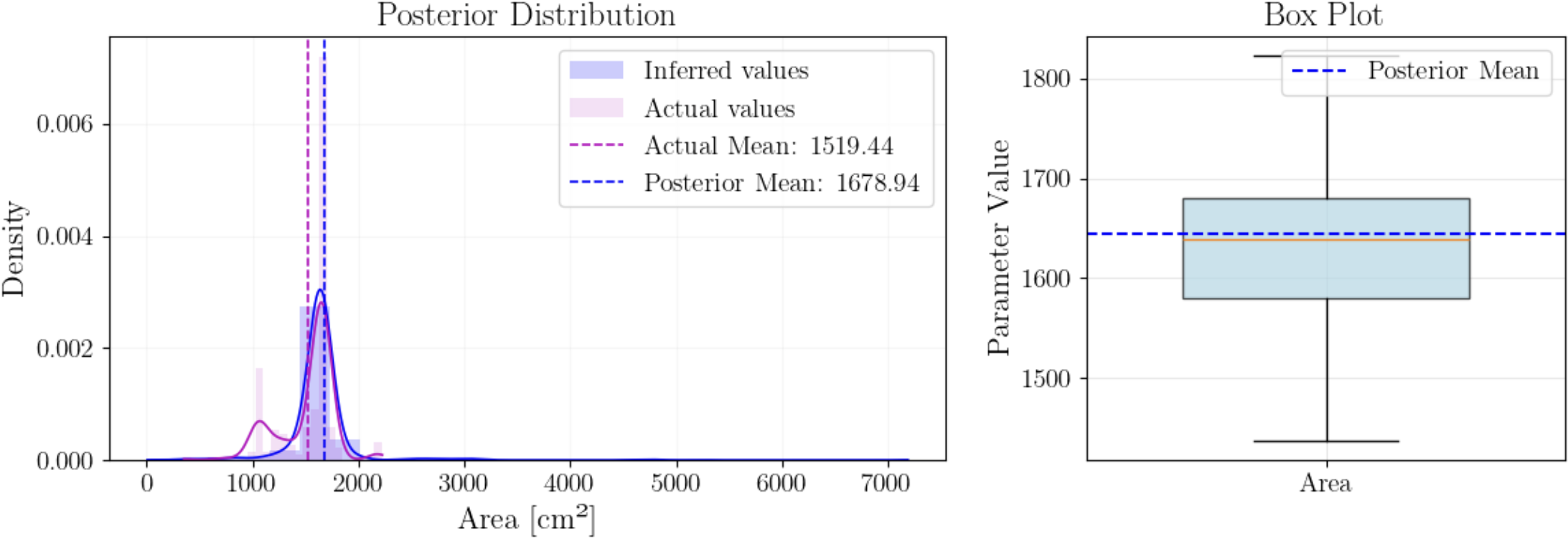}
				\caption{}
			\end{subfigure}
			\begin{subfigure}[b]{0.65\textwidth}
				\includegraphics[width=1\textwidth]{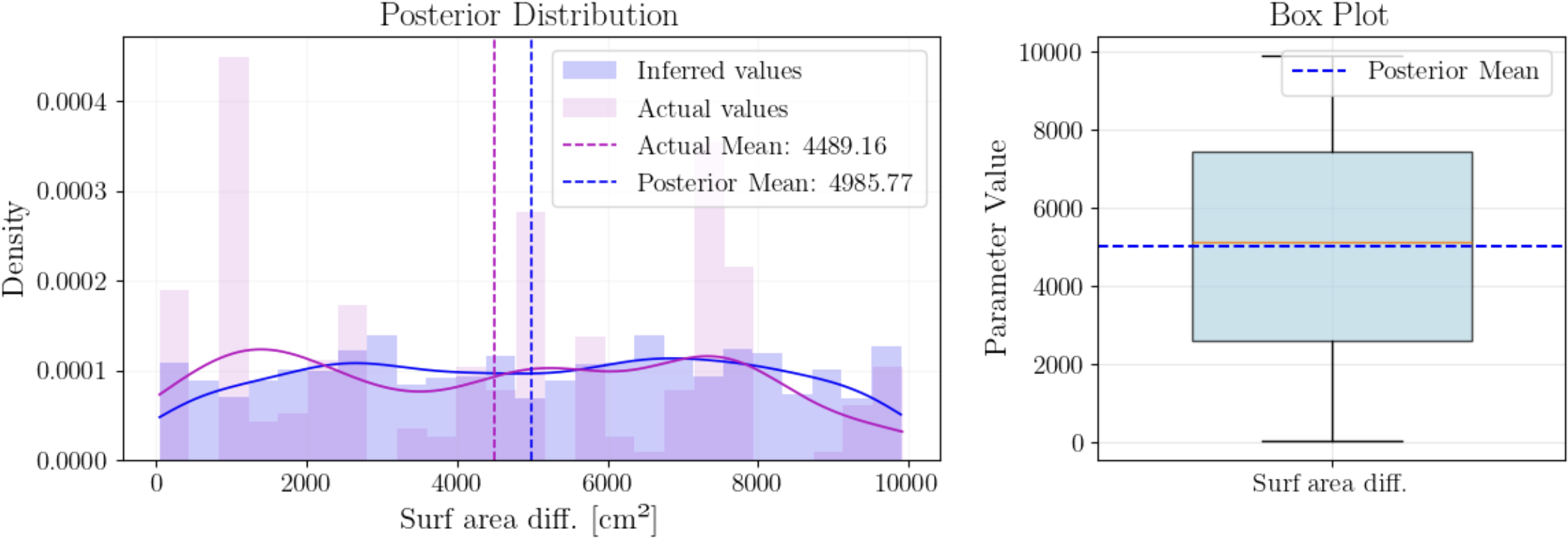}
				\caption{}
			\end{subfigure}
			\caption{For circular insert shapes: posterior distributions and credible intervals for key process parameters inferred using weighted ABC sampling (a) for "Pieces", (b) for "Area", and (c) for "Surface area difference". Left figures show kernel density estimates comparing inferred posterior distributions (blue) with actual parameter distributions from historical data (violet). Dashed vertical lines indicate their corresponding means. Right figures display boxplot-style credible intervals showing the 95\% CI (outer boundaries), 50\% CI (inner box, light blue), posterior median (orange line), and posterior mean (blue line).}
			\label{fig:Invcirc}
		\end{figure}
		\begin{figure}[!ht]
			\centering
			\includegraphics[width=.4\linewidth]{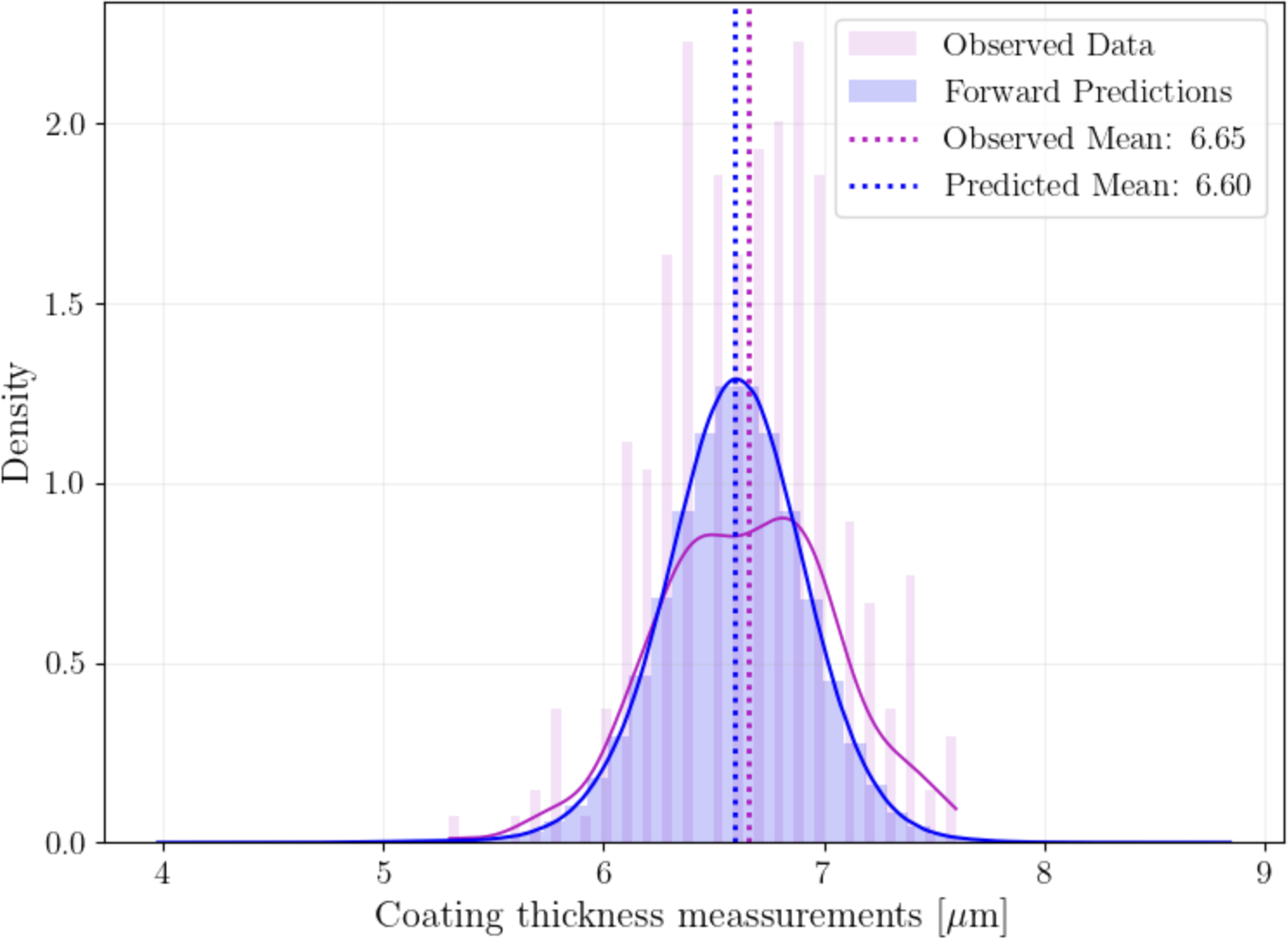}
			\caption{For circular insert shapes:: forward validation comparing observed process data with predictions generated from posterior parameter samples. Histograms and kernel density estimates show the distribution of insert thickness for observed data (orange) and forward predictions using ABC-inferred parameters (blue). Vertical dashed lines indicate distribution means.}
			\label{fig:outcirc}
		\end{figure}

		\newpage
		\subsection{Posterior distributions for rectangular insert shapes }\label{Appendix5}
		\begin{figure}[!ht]
			\centering
			\begin{subfigure}[b]{0.65\textwidth}
				\includegraphics[width=1\textwidth]{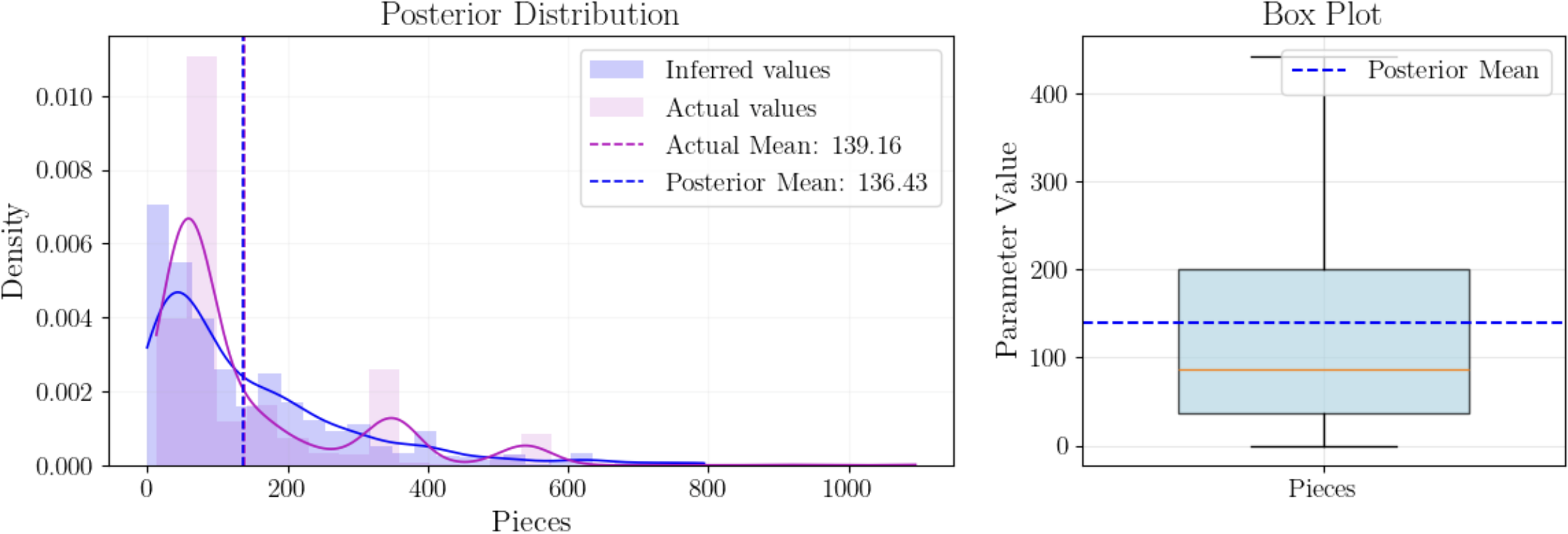}
				\caption{}
			\end{subfigure}
			\begin{subfigure}[b]{0.65\textwidth}
				\includegraphics[width=1\textwidth]{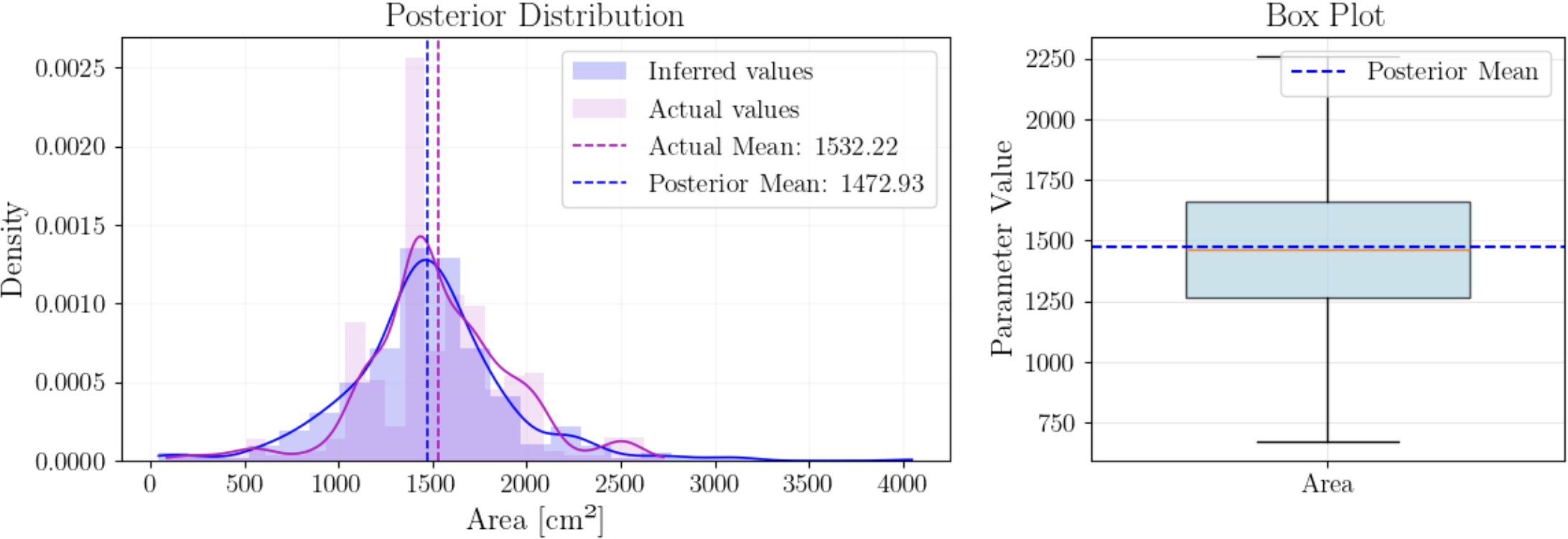}
				\caption{}
			\end{subfigure}
			\begin{subfigure}[b]{0.65\textwidth}
				\includegraphics[width=1\textwidth]{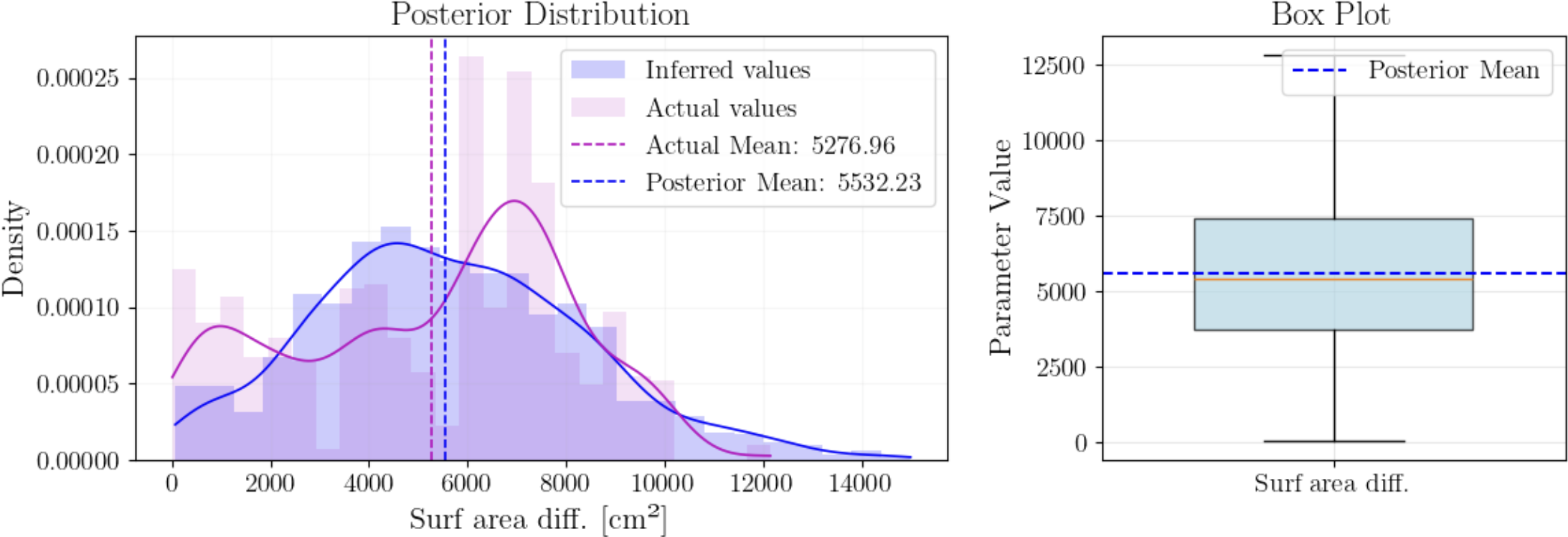}
				\caption{}
			\end{subfigure}
			\caption{For rectangular insert shapes: posterior distributions and credible intervals for key process parameters inferred using weighted ABC sampling (a) for "Pieces", (b) for "Area", and (c) for "Surface area difference". Left figures show kernel density estimates comparing inferred posterior distributions (blue) with actual parameter distributions from historical data (violet). Dashed vertical lines indicate their corresponding means. Right figures display boxplot-style credible intervals showing the 95\% CI (outer boundaries), 50\% CI (inner box, light blue), posterior median (orange line), and posterior mean (blue line).}
			\label{fig:Invrec}
		\end{figure}
		\begin{figure}[!ht]
			\centering
			\includegraphics[width=.4\linewidth]{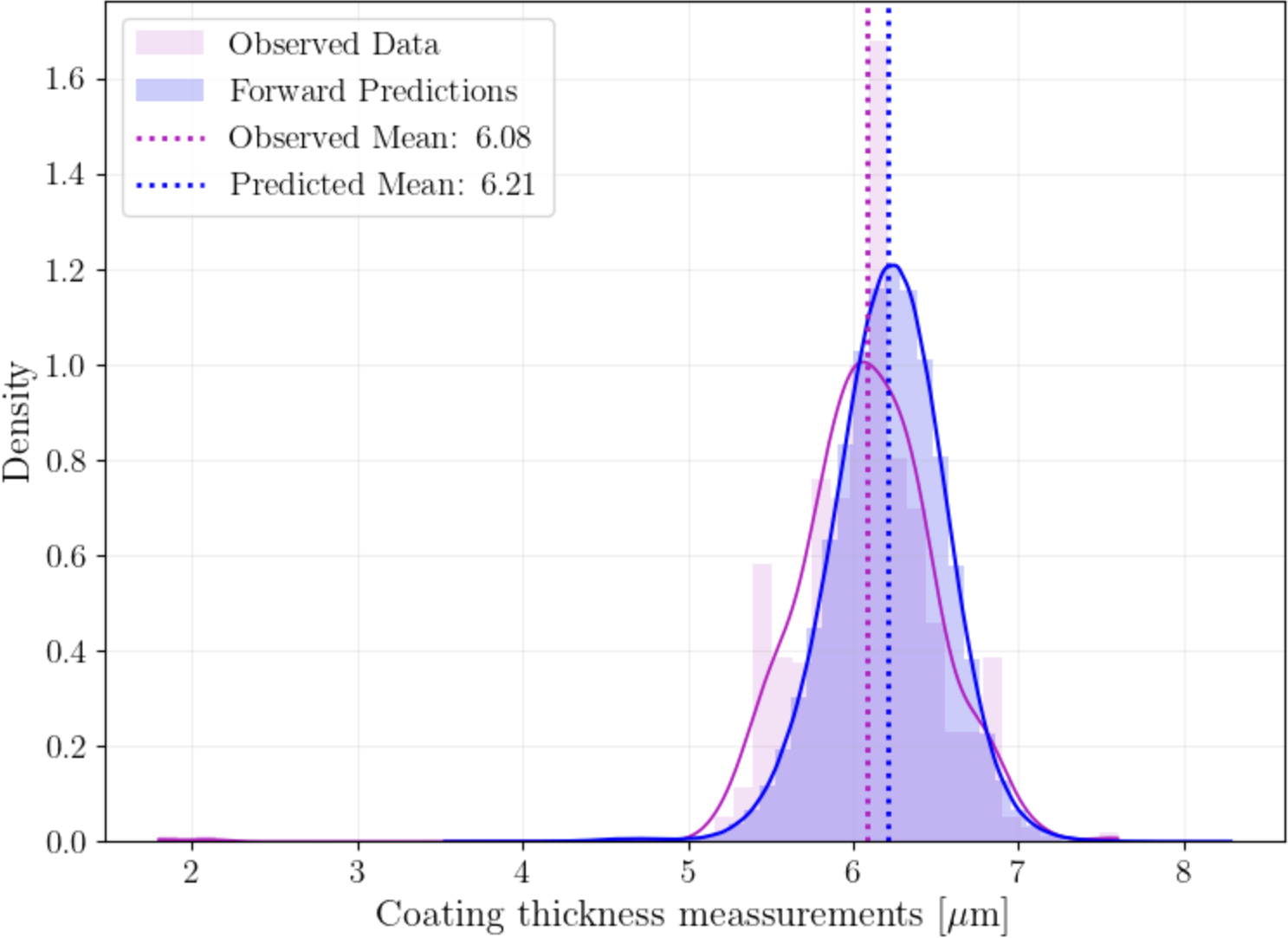}
			\caption{For rectangular insert shapes:: forward validation comparing observed process data with predictions generated from posterior parameter samples. Histograms and kernel density estimates show the distribution of insert thickness for observed data (orange) and forward predictions using ABC-inferred parameters (blue). Vertical dashed lines indicate distribution means.}
			\label{fig:outrec}
		\end{figure}
	\end{appendices}
	
\end{document}